# CVD diamond single crystals with NV centres: a review of material synthesis and technology for quantum sensing applications

J. Achard[1*], V. Jacques[2] and A. Tallaire[1,3]

[1] LSPM, Université Paris 13, Sorbonne Paris Cité, CNRS, 99, avenue JB Clément, 93430 Villetaneuse, France

[2] L2C, Laboratoire Charles Coulomb, Université de Montpellier and CNRS, 34095 Montpellier, France

[3] IRCP, Ecole Nationale Supérieure de Chimie de Paris, 11, rue Pierre et Marie Curie, 75005 Paris, France

* Corresponding author: jocelyn.achard@lspm.cnrs.fr

Abstract:

Diamond is a host for a wide variety of colour centres that have demonstrated outstanding optical and spin properties. Among them, the nitrogen-vacancy (NV) centre is by far the most investigated owing to its superior characteristics, which promise the development of highly sophisticated quantum devices, in particular for sensing applications. Nevertheless, harnessing the potential of these centres mainly relies on the availability of high quality and purity diamond single crystals that need to be specially designed and engineered for this purpose. The plasma assisted Chemical Vapour Deposition (CVD) has become a key enabling technology in this field of research. Nitrogen can indeed be directly *in-situ* doped into a high crystalline quality diamond matrix in a controlled way allowing the production of single isolated centres or ensembles that can potentially be integrated into a device.

In this paper we will provide an overview on the requirements for synthesizing "quantum-grade" diamond films by CVD. These include the reduction of impurities and surrounding spins that limit coherence times, the control of NV density in a wide range of concentrations as well as their spatial localization within the diamond. Enhancing the charge state and preferential orientation of the colour centres is also discussed. These improvements in material fabrication have contributed to position diamond as one of the most promising solid-state quantum system and the first industrial applications in sensing are just starting to emerge.

# 1. Introduction

Solid-state quantum systems that possess long-lived spin and/or optical coherence are likely to play key roles in the development of a broad range of applications in quantum technologies (QT), from quantum networks, to information processing and quantum sensing [1]. In this context, various candidates are being considered such as superconducting circuits [2], donors in silicon [3, 4], rare-earth ions in oxide crystals [5], quantum dots [6] or defects in semiconductors [7]. Among them, colour centres in diamond are arguably one of the most promising and studied systems [8]. The growing interest that surrounds this material essentially stems from the outstanding optical and spin properties of the nitrogen-vacancy (NV) colour centre [9], which have opened up a plethora of potential breakthrough applications in QTs, including the first demonstration of kilometer-scale entanglement between solid-state spin qubits for quantum networks [10], the realization of quantum error correction protocols [11], and the development of highly sensitive quantum sensors [12-14], which are already close to commercial products [15]. In addition, the progresses achieved in harnessing NV centres' potential for QTs has fostered the emergence of other colour centres in diamond, particularly those of group IV, that exhibit complementary properties such as SiV, GeV, SnV, or PbV [16-20]. **At the heart of the success of diamond as a platform for QTs, are the fundamental science and the technologies that have allowed the fabrication of specially designed and engineered *"quantum grade"* synthetic crystals.** Indeed, most practical demonstrations and advances in diamond-based QT leverage on material development.

Both the High-Pressure High-Temperature (HPHT) and Chemical Vapour Deposition (CVD) techniques are currently used to fabricate diamonds with optimized properties for industrial and high-tech applications. These synthetic diamond growth technologies have witnessed tremendous improvements over the past decades leading to ever thicker, larger and higher purity crystals [21]. Gem-quality material with fancy colours or colourless and up to several carats in size have been obtained, which may be seen by some, as a threat to the stability of the natural diamond market established for jewellery [22]. One of the driving forces for innovation however, has been the field of electronics in which diamond detectors as well as power devices are regarded as technologically disrupting with outstanding figures of merit [23]. Schottky diodes and field effect transistors are foreseen to allow the operation of smaller components that can drive exceptionally high currents and sustain high voltages in harsh environments [24-26]. While these devices are still in their infancy, their development has required improvements in the synthesis of fairly thick diamond films (several hundreds of µm) with a purity down to the ppb (parts per billion) level and a surface area as large as possible to facilitate processing and integration [27]. To this end, electrical doping using boron for p-type [28] and phosphorous for n-type [29] has been explored. Nitrogen which is a deep passivating donor with an activation energy of about 1.7 eV [30] generally needs to be avoided. However, like most wide band gap semiconductors, diamond suffers from a high activation energy of its dopants and asymmetric doping with n-type being extremely difficult to achieve on a standard (100) orientation and requiring non-conventional growth conditions and substrates [31-33]. **The enormous progresses made in this area during the last decades have played a crucial role in unleashing the potential of this material for QTs and have contributed to make diamond material available to this broad research community.** In fact, an accurate control over the amount of residual impurities (such as nitrogen and boron), isotopic content as well as crystalline defects that have a deleterious effect on spin coherence times are key to material adoption in QTs. While HPHT can produce bulk crystals with high crystalline perfection, purity remains limited and the technique is usually not flexible enough to allow for a precise engineering of "quantum grade" layers of material. The route that is thus most widely followed is to homoepitaxially grow a thin diamond film with optimized properties by CVD on a HPHT diamond substrate possessing appropriate crystalline quality and orientation. Although the CVD growth method is relatively mature, crystalline films that can deliver optimized performance in QTs are yet not routinely produced.

In this review we discuss some of the achievements and the remaining challenges that are crucial to the highly demanding field of diamond-based QTs, with a focus on quantum sensing and imaging applications with NV colour centres. In particular we give special emphasis to the material fabrication through the now well-established CVD technique and focus on *in-situ* doped material with NV centres for sensing devices. Issues with other colour centres (SiV, GeV etc.) might be raised but will not be discussed in great details.

## 2. Material requirements for NV-based quantum sensing applications

In this section, we fist identify some key challenges in diamond growth to optimize the performance of quantum sensing applications based on NV colour centres, starting with a brief reminder of their main optical and spin properties.

The NV colour centre consists of a substitutional nitrogen atom (N) combined with a vacancy (V) in a neighboring lattice site of the diamond crystal (figure 1(a)). This point-like defect gives rise to localized electronic states with energy levels deeply buried inside the bandgap of diamond. As a result, the NV centre can be considered as an *artificial atom,* mostly decoupled from the valence and conduction bands of the host material. Like many point defects in semiconductors, the NV colour centre can be found in various charge states having very different optical and spin properties [9]. Applications in QTs mostly rely on the negatively-charged state (NV$^-$), for which an additional electron is provided by a nearby donor impurity, thus leading to a quantum system with two unpaired electrons. The NV$^-$ colour centre exhibits a perfectly photostable photoluminescence (PL) emission with a zero-phonon line at 1.945 eV ($\lambda_{ZPL}$ = 637 nm), and provides a spin triplet ground level, which can be initialized by optical pumping, coherently manipulated with long coherence time through microwave excitation, and readout by pure optical means (figure 1(b)) [9]. As explained below, these properties are at the heart of NV$^-$ based quantum sensing. However, the NV defect can also be stabilized in a positively-charged configuration (NV$^+$), which is optically inactive [34, 35], and more often in a neutral form (NV$^0$), which is characterized by a shift of the zero-phonon line to 2.15 eV ($\lambda_{ZPL}$ = 575 nm), and does not feature the appealing spin properties of its negatively charged counterpart [36-38]. As a result, a first requirement on diamond crystals for QT applications is to provide an environment promoting the stabilization of the NV$^-$ charge state. In the following, we focus on the spin properties of the NV$^-$ configuration, which will be simply referred to as NV for clarity purpose.

A key feature of the NV colour centre is that its ground level is a spin triplet state, S = 1, whose degeneracy is lifted by spin-spin interaction into a singlet state of spin projection $m_s$=0 and a doublet $m_s$=±1, separated by 2.87 GHz in the absence of magnetic field (figure 1(b)). Here $m_s$ denotes the spin projection along the NV defect quantization axis, corresponding to a [111] crystal axis joining the nitrogen and the vacancy. Radiative transition selection rules associated with the spin state quantum number lead to an efficient polarization of the NV defect in the ground state spin level $m_s$=0 by optical pumping. Furthermore, the NV defect PL intensity is significantly higher when the $m_s$=0 state is populated. Such a spin-dependent PL response enables the detection of electron spin resonance (ESR) on a single defect by optical means. Indeed, when a single NV defect, initially prepared in the $m_s$=0 state through optical pumping, is driven to the $m_s$=±1 spin state by applying a resonant microwave field, a drop of the PL signal is observed, as depicted in figure 1(c).

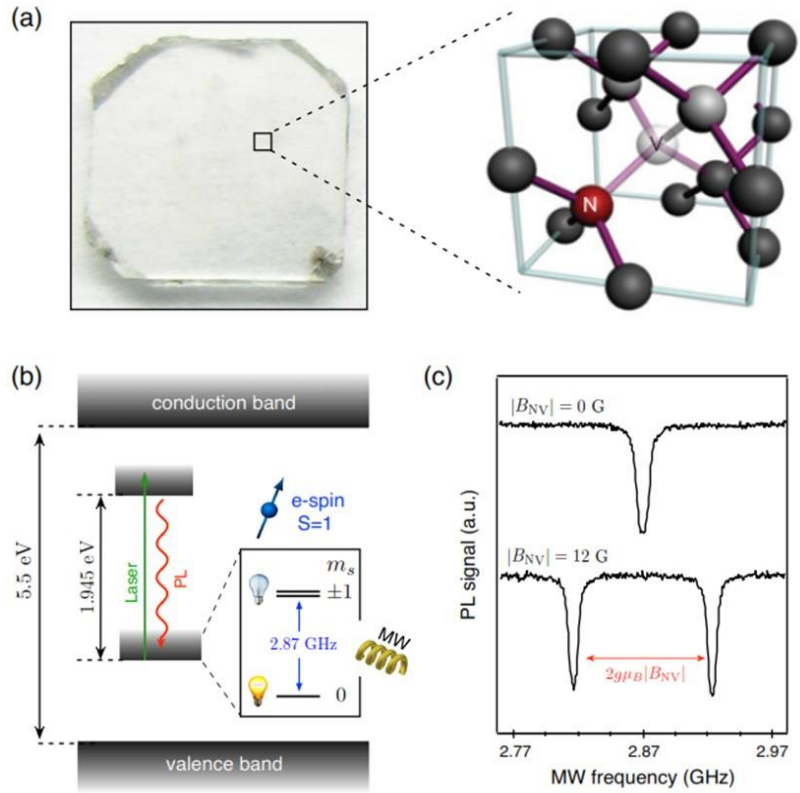

**Figure 1.** (a) Left panel: Optical image of a high-purity diamond crystal grown by CVD. Right panel: Atomic structure of the NV defect. (b) Simplified energy level scheme. The NV defect is polarized into the spin sublevel $m_s$=0 by optical pumping, and exhibits a spin-dependent photoluminescence (PL) intensity. (c) Optically-detected ESR spectra recorded by monitoring the NV defect PL intensity while sweeping the frequency of the microwave (MW) field. When a magnetic field is applied (lower panel), the ESR transitions are shifted owing to Zeeman effect thus providing a quantitative measurement of the magnetic field projection $B_{NV}$ along the NV defect quantization axis.

The first demonstration of optically-detected ESR on a *single* NV defect was reported in 1997, using a natural diamond sample and a confocal optical microscope operating under ambient conditions [39]. About ten years later, it was shown that these properties can be exploited for the design of a new generation of magnetometers [40-43], providing an unprecedented combination of spatial resolution and magnetic field sensitivity, even at room temperature. Here the magnetic field is evaluated within an atomic-sized detection volume by recording the Zeeman shift of the NV defect's electron spin sublevels (figure 1(c)), which is given by $\Delta = 2g\mu_B |B_{NV}|$, where $g\mu_B \approx 28$ GHz/T and $B_{NV}$ is the magnetic field projection along the NV defect quantization axis. The sensing functionalities of NV defects were then extended to a large number of external perturbations including strain [44], electric fields [45], pressure [46] and temperature [47-49], that all have a direct impact on the ESR frequency. For all these physical quantities, the shot-noise limited sensitivity $\eta_s$ of a single NV spin sensor scales as [13, 50]

$$\eta_s \propto \frac{1}{C_s\sqrt{RT_2^*}} \qquad (1)$$

where $C_s$ is the contrast of the optically-detected ESR spectrum, $T_2^*$ denotes the inhomogeneous spin dephasing time of the NV defect which limits the ESR linewidth, and $R$ is the number of detected photons. For a single NV defect, the ESR contrast is of the order of $C_s \approx 20\,\%$, a value fixed by the intrinsic photophysical properties of the NV defect, which can hardly be modified. The sensitivity can thus be improved either by increasing the collection efficiency of the PL signal [50] or by introducing alternative methods to improve the spin readout fidelity, such as photoelectric detection [51], spin-to-charge conversion [52] or infrared absorption readout [53]. From a material science point of view, the only parameter allowing to optimize the sensitivity is here the spin dephasing time $T_2^*$ of the NV sensor, which is mainly limited by magnetic interactions with a bath of paramagnetic impurities both inside the diamond matrix and on its surface [54]. A key requirement is therefore to engineer diamond samples with an extremely low content of impurities, as close as possible to a perfectly spin-free lattice, in order to reach long spin coherence times. We note that the sensitivity can also be enhanced for the measurement of time-varying signals. Such AC sensing protocols rely on dynamical decoupling sequences of the NV spin sensor, which results in a prolongation of its coherence time to a value commonly referred to as $T_2$, which can be orders of magnitude longer than $T_2^*$ [50].

While a single NV defect provides an ultimate spatial resolution for imaging applications, the sensitivity can be simply improved by increasing the number $N$ of sensing spins. For an ensemble of NV defects, the shot-noise limited sensitivity $\eta_e$ then scales as

$$\eta_e \propto \frac{1}{C_e \sqrt{NRT_2^*}} \quad (2)$$

A challenge in material science is thus to increase the density of NV defects while maintaining good spin coherence properties. However, the gain in sensitivity is partially compensated by a reduced contrast of spin readout. Indeed, NV defects are oriented with equal probability along the four equivalent <111> crystal directions, leading to a decreased sensitivity because only a quarter of NV spins are efficiently contributing to the detected signal, the others producing solely a background photoluminescence. In addition, luminescence from other impurities, such as the neutral $NV^0$ defects, further impairs the signal to background ratio. The spin readout contrast then falls typically to $C_e \approx 1\,\%$ for large ensembles of NV defects [50]. Mitigating this effect requires (i) to achieve preferential orientation of the NV defects during the diamond growth and (ii) to improve the conversion of NV defects in the negatively-charged state configuration.

Besides providing the highest sensitivity to date [55], ensembles of NV defects can also be used for imaging applications [56, 57]. To this end, a sample of interest is commonly deposited directly on top of a diamond crystal, which contains a thin layer of NV centres near the diamond surface. The spin-dependent PL signal from the NV layer is imaged onto a CCD camera in a wide-field detection scheme, with a spatial resolution limited by diffraction (∼500 nm). In the last years, this method has found numerous groundbreaking applications in very different fields of research [12], including NMR spectroscopy [58, 59], biomagnetism [60], geoscience [61], and condensed matter physics [62-65]. Further performance improvements of this technique require to engineer thin diamond layers with a high NV density featuring long spin coherence time and preferential orientation.

To summarize, current challenges in diamond growth to optimize the performance of NV-based quantum sensing include

(i) **Tailoring the diamond matrix so that decoherence is as limited as possible.** Although important progresses have been obtained through defect engineering or isotopic purification, coherence times are still far from the theoretical $T_1$ limit. This is particularly true when the spins of interest are located near the surface or in a diamond crystal with a high nitrogen content.
(ii) **Controlling NV density and charge state.** Since many sensing applications rely on dense NV ensembles to improve sensitivity, controlling the ratio between NVs and other N-containing defects is crucial. In addition, the close environment of the defect has to favour the occurrence of the negative charge state with respect to the neutral one.
(iii) **Spatially localizing NV centres.** The precise positioning of single or ensemble of NV centres both in-depth and in-plane is of importance for incorporating them into cavities or nanostructures, or for improving the performances of wide-field imaging with NV ensembles.

(iv) **Controlling NV orientation**. Promoting preferential orientation is desirable to limit background noise level, increase sensitivity and simplify device operation.

These different aspects about the material fabrication will be discussed hereafter.

## 3. The synthesis of "quantum grade" diamond films and crystals
### 3.1. HPHT grown diamonds

HPHT is well established to produce bulk single crystals with up to a few millimetre thick and millimetre square size that are available commercially in particular for cutting tools applications. This technique typically uses a bath of melted transition metals (such as Co, Fe, Ni, Cr, Mn etc.) in which carbon (in general diamond powder or graphite) is dissolved and re-precipitated on the facet of a small seed in a region of slightly lower temperature (20 to 50 °C less) [66]. This temperature gradient approach involves pressures and temperatures above 5 GPa and 1300 °C respectively. Different heavy set-ups exist that differ from the way the pressure is applied to the cell, such as uniaxial compression with belt and toroid systems, or multi-anvil systems (so-called bars and cubic presses) [67]. This equipment is mostly operated by industrial players (*Element Six, General Electrics, Sumitomo, New Diamond Technology* etc.) and are essentially destined to mechanical applications for which requirements on purity and quality are moderate. Under adapted and stable conditions though, large crystals can be produced with potentially extremely low extended defect content. For instance inclusion-free single sector diamonds with stacking faults and dislocation content below a few hundred per cm² have been demonstrated [68, 69]. However the precise recipes developed to reach this degree of perfection are usually a well-kept industrial secret, while the associated costs can be tremendously high (up to several k€ for a 500 µm thin slab). In contrast, standard HPHT crystals typically contain dislocation densities of the order of $10^4$-$10^5$ cm$^{-2}$ and visible growth sectoring [70], but their cost is limited to a few hundreds of euros depending on size, orientation and polishing.

Although bulk crystals with low dislocation density can be grown by HPHT, the technique is not well adapted to produce films with a high purity or a controlled doping as required by quantum applications. Indeed, the starting powdered materials and the high pressures needed promote incorporation of impurities that may come from surface contaminations or are trapped in porosity. Standard HPHT diamonds are usually labelled as type *Ib* due to the presence of a large amount of non-intentionally doped nitrogen in them (typically 10-300 ppm) that lead to a yellowish colouration and obvious growth sectoring (see figure 2(a)) [71]. Nitrogen uptake depends on the solvent-catalyst used and its solubility in them. While quantum sensing requires incorporation of NV centres, it should be emphasized that nitrogen content in HPHT crystals is not necessarily in the form needed for an optimized sensor. A significant fraction of N is for example present as substitutional ($N_s$), known as P1 centre. Aggregated forms also exist due to nitrogen mobility being activated under high pressures and temperatures. The N-V-N or H3 centre is for example frequently created and leads to emission at 503 nm in PL. Other aggregated forms are less prevalent but commonly found like A-centres (2 neighbouring $N_s$) or B-centres ($N_4V$) [71]. Their concentration can be of the order of several ppm depending on the growth conditions or treatment that they have underwent. The addition of getters (Ti, Zr, Al etc.) can reduce the amount of incorporated nitrogen by preferentially associating and precipitating it as a nitride allowing fabrication of type *IIa* colourless diamond crystals by HPHT (see figure 2(b)) [72]. This material leads to lower background PL and narrower diamond Raman peaks as illustrated in figure 2(c). However in general, N content cannot be suppressed completely and remains of the order of 0.1 ppm. Growth rates under such low nitrogen conditions are also strongly reduced which increases the overall cost of the HPHT diamonds [69].

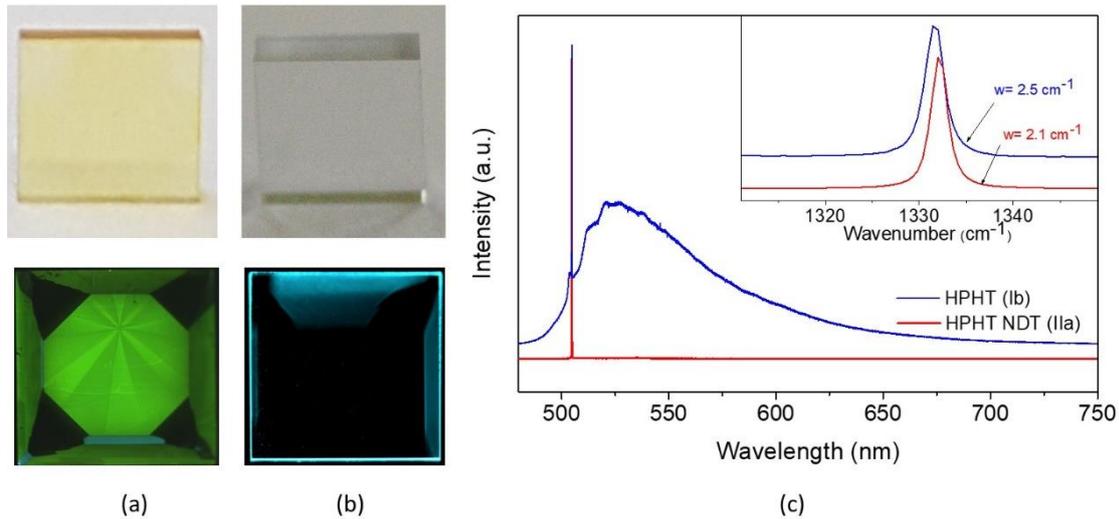

**Figure 2**. (a) Optical and PL images of a type *Ib* HPHT substrate. (b) Optical and PL images of a type *IIa* HPHT substrate. (c) Raman/PL spectra of type *Ib* and *IIa* HPHT substrates obtained with an excitation line at 473 nm. The inset shows the linewidth of the diamond Raman peak. The type IIa diamond was provided by New Diamond Technology (NDT).

Other impurities may also be incorporated in significant amounts including solvents from the melted baths (Ni and Co) or element contaminations (B, Si, Ge) which can lead to the appearance of specific defects or colouring [73]. Around 50 ppm of boron have been measured in some of the purest type *IIa* diamonds [74]. Boron is known to stabilize the neutral charge state of NV centres and is usually not desirable (see part 4.4). Impurity incorporation dependence on crystal orientation is also an important issue that is associated with variations in colouring and/or luminescence under UV light (figure 2(a)). Incorporation efficiency in (111) growth sectors is usually 2-3 times higher than in (100) and (110) sectors [75]. Isotopic purification of HPHT diamonds to change the $^{12}C/^{13}C$ ratio has been achieved using pyrolytic carbon powder [76] but it remains relatively difficult and uncommon due to the high cost of the precursors and low flexibility of the technique. While residual impurities are difficult to avoid with this synthesis process, intentional additions of certain metals to the bath/catalyst mixture can be explored to create specific colour centres. For example, SiV, GeV or SnV centres which are also interesting systems for QTs have been obtained [77, 78]. This is an important advantage of the HPHT approach because such elements cannot always be easily brought in through the vapour phase or incorporated using the CVD production technique due to limited solubility.

The ability to control crystal morphologies through tuning of the growth temperature and the solvent has been highlighted and opens the way to obtaining various crystal habits from cubic to octahedral. Control of the morphology opens the way to the fabrication of larger plates with specific orientations that can be extracted from such stones [79]. In particular [111]-oriented diamonds can be cleaved from octahedral shape crystals and are particularly suited as substrates for CVD overgrowth with oriented NV centres (as will be discussed in part 6) [80]. Bulk crystals can thus be obtained through the HPHT technique with a high crystalline perfection but limited purity. Nevertheless NV ensembles in type *Ib* HPHT diamonds have been studied and exploited for QT demonstrations. Some examples include magnetometry [55], MW photon storage [81], coupling to superconducting resonators [82], quantum memories [83], hyperpolarisation of $^{13}C$ [84] or data storage [85]. Although one can benefit from a bulk material that is easily available, the nitrogen density in the form of substitutional defects is usually a limiting factor and reduces the coherence times ($T_2$) to typically 1-2 µs only at room temperature [86]. Nevertheless, by reducing the spin bath surrounding NV centres through isotopic purification and limited nitrogen doping, as well as irradiating the crystal to convert $N_s$ into NV, $T_2$ can be extended to several tens of µs [81]. In general electron irradiation followed by annealing has become a rather standard treatment to improve the performance of such HPHT crystals (see part 4.1). Although some attempts have been made to explore bulk HPHT diamonds crystals in the field of quantum sensing, this material in general fails to provide the purity and the flexibility of fabrication that is required for highly efficient devices. The most common approach thus relies on using HPHT crystals as the starting substrates onto which films with the desired properties are overgrown by CVD.

### 3.2. CVD grown diamonds

Microwave plasma assisted CVD has become a key technology showing great potential to produce engineered films with the desired doping, isotopic purity and dimensions. The use of CVD-grown diamond films and crystals is relatively wide-spread in QTs although their availability remains scarce. Unlike HPHT, the CVD technique mostly involves academic research groups while commercial availability of high purity plates is limited to a few industrial companies only (*Element Six, Diamond Materials, IIa Technologies* etc.). A large market for CVD diamond plates is yet to be found. The fabrication of high-quality thick crystals is also technologically challenging with difficult scaling-up which contributes to increase the fabrication costs. Nowadays, HPHT still remains dominant when it comes to producing bulk synthetic diamonds while CVD is mostly focused on producing thinner layers. Although the size and thickness of the produced crystals is not such a limiting factor for QTs, diamonds may need to be thick enough to be processed and properly oriented or separated from their substrate.

The CVD technique operates at low pressures (10-300 mbar), under conditions at which graphite should be the thermodynamically stable phase [87-89]. It involves kinetically stabilizing diamond through the production of atomic hydrogen within a high temperature plasma media that preferentially etches away weak $sp^2$ bounds, allowing the addition of carbon to the diamond lattice of the substrate. $H_2$ and $CH_4$ are used in a typical proportion of 95-99 % to 5-1 % respectively. Addition of $O_2$ in a small amount (< 2%) is sometimes used in order to increase the etching effect and limit impurities incorporation or non-epitaxial defects formation [90, 91]. In general, activation of the gas is performed through applying a 2.45 GHz MW field to a resonant cavity reactor (figure 3(a)) [92, 93]. Operation under higher pressures (> 100 mbar) and microwave powers (> 2 kW) leads to the formation of a localized plasma region in the core of which temperatures may reach up to 3000 K which is favourable to produce precursors for growth [94]. Indeed, thermal dissociation of the molecules into a variety of atomic and radical species is highly pronounced and may be of up to several tens of percent [95]. Growth is carried out at a temperature in the range 700-1100°C on a diamond seed through either cooling or heating the substrate holder depending on the power density applied. Several providers commercialize MW plasma assisted systems with varying characteristics (*Cornes Technologies (Seki systems), Plassys, iplas, optosystems* etc.) but a large number of research groups have developed their own equipment. The main differences in those systems are essentially in the way the MW radiation is coupled to the resonant cavity (electromagnetic modes), the location of dielectric windows as well as the design of the holder (translatable, rotatable, cooled or heated). High-power operating reactors are preferred for achieving high growth rates and low defect bulk diamond crystals. However the low-power regime may be advantageous for ensuring nm-scale control over the thickness of the layers and a precise positioning of dopants or colour centres (see part 5.1).

With a hetero-substrate (like a silicon wafer) a polycrystalline film in which grain size directly depends on thickness through a columnar growth mode is generally obtained. Under certain growth conditions, films may exhibit a particular texture or preferential orientation [96]. The presence of grain boundaries is however deleterious to obtaining long coherence times and low background luminescence. Polycrystalline diamonds are usually not preferred for sensing applications. Interesting properties for single NVs have however locally been found within the larger grains of polycrystalline films [44, 97]. They also offer the advantage of providing a large and flexible platform for processing them into photonic crystals and resonators that would be highly desirable for QTs [98].

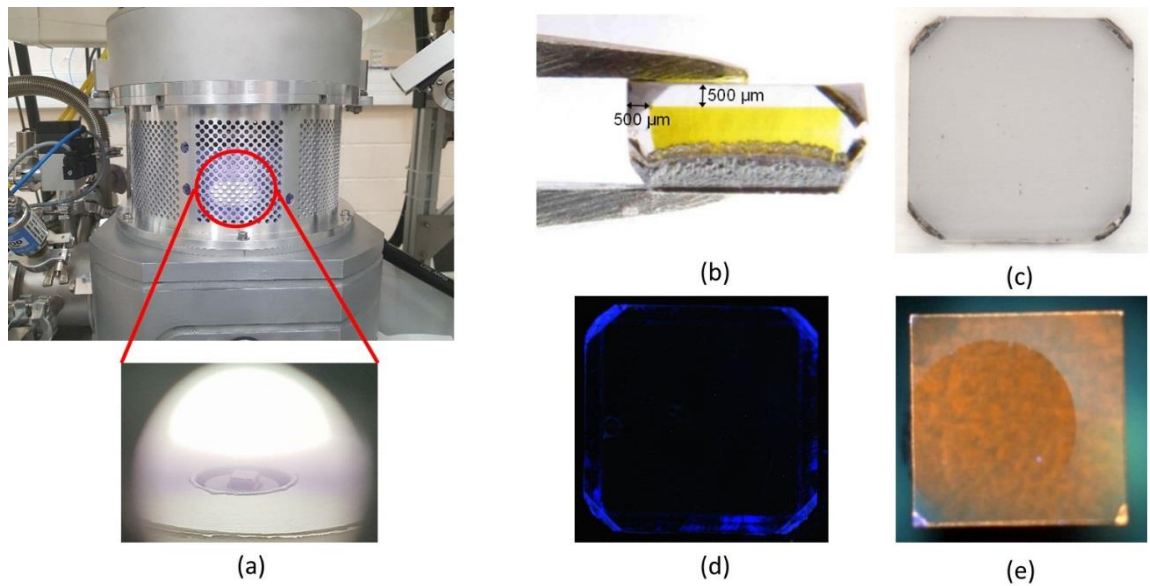

**Figure 3.** (a) Microwave Plasma Assisted CVD system allowing the growth of diamond. The inset shows a zoom into the plasma region in which a diamond is positioned for growth. (b) High purity thick CVD diamond layer grown on a yellow HPHT diamond substrate and (c) freestanding CVD diamond film obtained after removing the HPHT substrate by laser cutting and polishing. (d) PL image of a high purity thick freestanding CVD diamond film showing only very weak blue luminescence in the corners corresponding to the presence of stress. (e) PL image of a $N_2$ doped thick freestanding CVD diamond film showing orange luminescence corresponding to the presence of both $NV^0$ and $NV^-$ colour centres.

Heteroepitaxial growth on specially developed templates that include a thin monocrystalline iridium layer deposited on an oxide thin film on silicon or a bulk crystal (Yttria stabilized $ZrO_2$, a-plane $Al_2O_3$ or $SrTiO_3$) is another possible approach that promises wafer-scale deposition area [99-101]. In recent years, some important efforts in material development have been devoted to obtaining higher quality and larger films with new venture companies starting to commercialize them (*Audiatech, Namiki* etc.). The complex steps that lead to oriented diamond growth include deposition of the epitaxial Ir films on an appropriate substrate, the biased enhanced nucleation of diamond domains on them, thickening of the film and limitation of dislocation density through patterning of the surface [102]. In general dislocation densities still remain high even in thick films (> $10^7$ cm$^{-2}$) [103] and impurities such as silicon are hard to avoid. Nevertheless a recent assessment of state-of-the-art films produced through this approach have demonstrated coherence times of 5 µs supporting the idea that they may provide a useful larger platform for future applications in QTs providing material quality and availability are improved [104].

Homoepitaxial growth onto a diamond seed (generally a type *Ib* HPHT substrate) is the preferred fabrication route for obtaining high quality and purity diamond films that are suitable for QTs (figure 3(b) and 3(c)). Although CVD is relatively simple in its operating principle, obtaining layers with a given defect concentration and the desired thickness has animated a large number of research activities through the past decades, in particular within the earlier and demanding context of power electronics. Substrate selection and preparation plays an important role in the epitaxial overgrowth and adapted polishing or etching of the surface prior to growth helps limiting the propagation of defects from the interface [105]. Maintaining constant growth conditions, especially temperature, during long periods of time is also a limiting factor when thicker layers are desired. The development of specific substrate holders that include cooling with gas mixtures or vertical translation may be needed [106-108]. The presence of uncontrolled amounts of $N_2$ or $O_2$ from reactor leaks or impure feed gases have important consequences and can induce the formation of polycrystalline defects that would quickly ruin the entire growth run [87, 94, 109, 110]. To this end, care must be taken to frequently check for potential leakage sources and to use dedicated purifying systems especially for hydrogen. When a good control of gas environment is successfully achieved, single crystal diamond plates with good purity or intentionally doped with a controlled nitrogen amount can be prepared (see figure 3(d) and 3(e)). Other potential contamination sources may be released by the constitutive materials of the reactor themselves (metal walls, quartz windows, molybdenum holders etc.) such as boron, silicon or nitrogen. Besides choosing adapted materials for the reactor furniture, the design should ensure that internal parts are appropriately cooled down or positioned far away from the high temperature plasma media. The appearance of SiV emission in CVD diamonds is nevertheless very common and is even used as a criteria for establishing diamond's synthetic origin in gemmology [111]. Finally it should be noted that hydrogen, one of the main elements involved in the growth process is usually overlooked although it is one of the main impurities in CVD-grown crystals. Hydrogen-vacancy defects known as H1 centres are paramagnetic and show-up in EPR together with the nitrogen-vacancy-hydrogen (NVH) for example [112, 113].

The ability to prepare isotopically enriched layers is also a particularly useful asset of CVD grown diamond films. Indeed growth from methane using natural isotopic carbon ratio leads to the presence of 1.1 % of $^{13}C$ in the films which is a non-zero nuclear spin element. Coupling of the NV spins to nearby $^{13}C$ atoms is the main source of decoherence for films with a low NV amount (< 0.1 ppm) [54, 114]. Reducing the amount of $^{13}C$ is relatively straightforward by substituting the conventional methane source with an enriched $^{12}C$ methane cylinder. By doing so, $T_2$ times have been successfully extended from a typical value of 0.5 ms up to a record of 2.4 ms [115]. Nevertheless the cost of such sources is several orders of magnitude higher than a standard methane cylinder. Moreover the specifications in terms of $N_2$ or $CO_2$ background content are usually much higher than high-purity grade methane and may require additional purification steps with dedicated purifier cartridges. On the other hand, intentional addition of $^{13}C$ in CVD-grown films can be achieved to deviate from the natural isotopic ratio. Particular schemes have been proposed that explore coupling of a NV spin to a nearby long-lived nuclear spin to further extend quantum storage times [116]. Dynamic nuclear polarization may also be useful to magnetic resonance spectroscopy and imaging applications [117].

In general, one of the main advantages of the CVD growth approach for making "quantum grade" diamonds is the ability to engineer stacked layers of different doping and composition in a dynamic and very flexible way. Indeed the gas phase environment can be controlled to an extremely high level while changing from one composition to another can be done with abrupt interfaces providing residence time of gas species are taken into account (see part 5.1). CVD diamond fabrication of specially designed bulk crystals or thin films has thus become a cornerstone of the developments that the quantum technologies based on this material system have witnessed.

## 4. Creating colour centres with good coherence properties
### 4.1. Implanting colour centres in high purity CVD diamonds

While the CVD technique allows the fabrication of isotopically enriched diamond films with extreme purity, a varying amount of colour centres need to be incorporated within this matrix to provide the sensing functionality. A widely followed approach consists of locally implanting nitrogen ions ($N^+$) or a molecule containing nitrogen ($N_2^+$, $CN^+$ etc.) in "electronic grade" (i.e. high-purity and non-luminescent) CVD diamonds. In general 3 steps are required: (i) introducing impurities, (ii) creating vacancies (that may be co-implanted together with the impurity or afterwards), (iii) annealing to heal defects and diffuse vacancies so that the complex defect can be formed. The present paper does not intend to give a detailed review of the optimization of implanted colour centres in diamond and readers are advised to refer to the following articles [118, 119], however some of the main trends are presented below.

Regarding the first step, the ions to be implanted can be accelerated in a wide range of energies from typically 2 keV to 20 MeV leading to implantation depths of 3 nm to about 5 µm respectively. This obviously requires rather different implantation set-ups from small table-top sources for low energies to large tandem accelerators to reach the MeV regime. It allows creating specific luminescent patterns within the diamond substrate as illustrated in figure 4(a). It should be noted that the ion energy not only influences the penetration depth of the ions but also the creation yield of NV centres with respect to each N atom entering the diamond lattice [120]. In fact, the higher the energy, the higher the number of vacancies that are co-created leading to higher yields. Typically, values range from 0.1 % at 2 keV up to about 45 % at 18 MeV. However at high acceleration energies, spatially positioning the implanted ions with accuracy becomes difficult due to the statistic distribution of collisions with atoms in the lattice that leads to a lateral and depth spread called straggling. Getting a spatial accuracy of less than 5 nm for example requires that ions are accelerated to an energy below 10 keV which limits penetration to 10 nm only (i.e. to near-surface NV centres). This shows that a trade-off exists between high-yield high-depth NVs and low-yield low-depth but highly localized NVs, depending on the energy of the incoming ions [121]. To go beyond those limits, strategies have been developed to increase the positioning accuracy by implantation through a pierced AFM tip [122], mica channels or opened PMMA masks [123].

An additional advantage of the implantation technique relies on its ability to generate defects from elements that cannot be easily grown-in directly by CVD due to too high steric hindrance, low stability or difficulty in bringing them through the gas phase. Besides, co-implantation with other elements brings additional flexibility in the generation of complex defects. Lühmann et al. have for example studied a wide variety of colour centres that can be created through implantation of elements as varied as Mg, Ca, F, O etc. in a matrix that already contains other implanted impurities of phosphorous or boron [124]. This obviously leads to an exhaustive variety of combinations and adds additional complexity to this approach in determining the most relevant colour centre for a given application. Control of the charge state of created vacancies has been accomplished by implanting nitrogen into n-type doped diamond rather than in a standard intrinsic crystal. In this way, the negative charge state of the vacancies is promoted which reduces their clustering and thus increases the probability that they associate to a single nitrogen atom to form a NV centre. Record creation yields of about 75 % have been reported in sulphur doped diamonds [125].

Introducing additional vacancies into the diamond crystal can also be explored in order to boost the NV/Ns ratio (creation yield). Such irradiations are accompanied by the creation of the GR1 luminescent defect (neutral vacancies) which intensity depends on the dose, the energy and the type of ions that are used (figure 4(b)). Helium ions accelerated to a few keV provide for example a way to locally create vacancies at a controlled depth of a few tens of nm and thus the ability to generate so called delta-profiles [126]. He$^+$ ion beams can also be focused down to a small size to create patterns [127]. However, evidence exist of the creation of specific colour centres related to the implantation of helium atoms within the lattice. The optical properties of such Helium-Vacancy (He-V) centres have recently been studied [128]. Other irradiations using protons or electrons present the advantage of having a lower mass as compared to helium. Their stopping range is much longer which allows for a more uniform creation of vacancies through the volume of the sample. Electron irradiation at several MeV and doses of the order to $10^{17}$-$10^{19}$ cm$^{-2}$ has become a standard treatment to *Ib* HPHT diamonds in order to increase NV density [129]. Local irradiation at lower energies (around 200 keV) has also proved successful using the electron beam of a Transmission Electron Microscope (TEM) [130]. At this energy the penetration depth can be estimated to about 140 µm. Vacancy creation efficiency is also more limited with a minimum energy for vacancies creation of about 145 keV. Nevertheless this technique provides a way to generate local NV patterns [131, 132]. An alternative approach is the creation of vacancies through ultrafast (fs) laser irradiation pulses. Single NV centres can thus be written locally into arrays with a positioning accuracy of about 200 nm and coherence times of several hundreds of µs equivalent to naturally occurring NVs [133, 134].

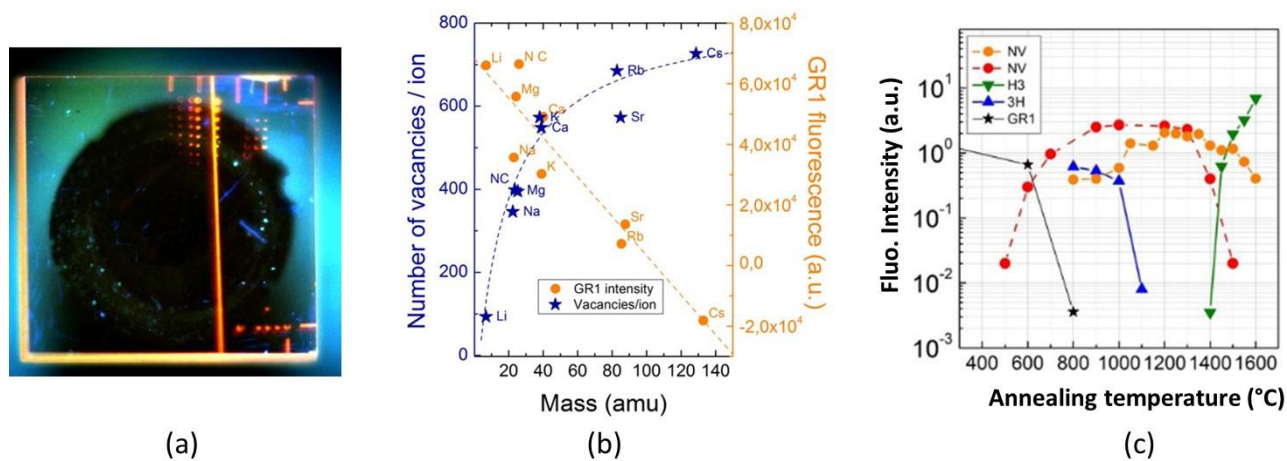

**Figure 4.** PL image of a high purity freestanding CVD diamond film after localized nitrogen ion implantation which leads to the appearance of orange/red spots corresponding to $NV^0$ and $NV^-$ colour centres (Collaboration University of Leipzig). (b) Evolution of vacancies/ion number and GR1 fluorescence as a function of implanted ion mass. (c) Evolution of the fluorescence of defects created by implantation as a function of annealing temperature. (b) and (c) are adapted from Lühmann et al [124].

Annealing ion implanted diamonds is a key ingredient to increase NV density and improve their coherence properties. This step allows vacancies to diffuse and defects to heal so that highly coherent NV centres are formed successfully following irradiation. Since vacancies in diamond are mobile above 700 °C, typical annealing temperatures after irradiation are in the range 800-1000 °C with the treatment carried out for a few hours (1-10 h). This is illustrated in figure 4(c) where the number of NVs is seen to increase with temperature together with a decrease of GR1. Annealing simultaneously when doing the ion implantation provides a way to reduce collateral damage and preferentially associate the vacancies with a nearby N rather than forming clusters [135]. Annealing at too high a temperature (> 1200°C) is likely to lead to the formation of vacancy clusters or to the thermal dissociation of NVs which should be avoided. Nitrogen atoms can also become mobile at temperatures of the order of 1600 °C possibly forming complex clusters like H3 (N-V-N) as shown in figure 4(c). In general $T_2$ times of implanted and annealed NVs remain below those of naturally created ones by 1-2 orders of magnitude (typically 1-10 µs) due to the presence of other defects and residual damage that cannot be completely annealed out. Optimized annealing treatments at higher temperatures [136, 137] or composed of successive steps with various annealing temperatures and durations have been proposed to obtain NV coherence times close to those of native NVs [138]. Nevertheless there probably does not exist a universal efficient annealing step, the optimized procedure strongly depending on the initial quality of the diamond as well as the starting NV density.

### 4.2. In situ doping of colour centres

While ex-situ creation of colour centres is a flexible approach with accurate positioning ability, it generally does not allow obtaining defects with as good coherent properties as naturally occurring ones from in-situ doping. Intentional doping during CVD growth can indeed be achieved by injecting into the plasma a precursor gas containing the element to be doped. $N_2$ is the most widely used dopant for NV doping. This molecule has a very strong bond energy (9.8 eV) which requires high plasma power densities to efficiently dissociate it. Nitrogen doping into the tight diamond lattice is also not energetically favourable. Both those issues lead to low doping efficiencies of about $10^{-5}$ to $10^{-3}$ [139, 140]. When low plasma power densities are used (< 1 kW MW power), several % of $N_2$ are usually required to reach $N_s$ concentrations of the order of 0.1 ppm [141, 142].

An additional limitation is that only a small fraction of the total nitrogen will be incorporated as a complex associated to a vacancy, with the main part being substitutional to a single carbon atom ($N_s$). A typical yield ($NV/N_{total}$) for untreated as-grown CVD diamonds is of the order 1/300 or below [143]. Therefore it should be highlighted that the creation yield for in situ doping is similar to that obtained for nitrogen implanted at medium energies. This is a particularly limiting factor since not only the amount of NV centres to be used for sensing will be limited but the large proportion of nitrogen, a paramagnetic impurity, will induce decoherence especially for the highest doping levels [114]. In addition a small part of the total nitrogen in the CVD-grown diamond, more or less the same proportion as that of NVs may occur in the form of NVH complex [144]. These defects are fairly common in CVD diamonds grown under high nitrogen additions since hydrogen is one of the main ingredients for CVD growth. They can be detected in their negative form as a line in EPR and are also visible as a sharp absorption at 3123 cm$^{-1}$ in FTIR in their neutral charge state [145, 146] (figure 6(c)). While hydrogen impurities are likely to passivate part of the NV centres and produce additional magnetic noise, these complexes cannot be easily annealed out even at very high temperatures. It would be desirable to improve the NV yield by changing the growth conditions such as substrate orientation, pressure and MW power, gas phase composition (methane, hydrogen, nitrogen and oxygen). However no systematic study exists so far on the influence of growth conditions on the NV creation yield most likely due to the difficulty in accurately measuring the concentration of those defects in thin diamond films.

While $N_2$ is the most frequently used dopant, other molecules have been shown to lead to improved NV doping efficiency and photostability. For instance, $N_2O$ which has a much lower dissociation energy is also available as a high-purity gas. While high-density NV ensembles (around 10 ppb) created through addition of $N_2$ are subject to blinking and charge state instability especially under high laser pumping power (figure 5(a-c)), those formed from $N_2O$ are much more stable (figure 5(d-f)) [147]. This is possibly related to the presence of a low amount of oxygen near the growing surface due to $N_2O$ dissociation in the plasma that etches away any defects that act as traps for charge carriers [148]. However the use of other dopant sources is not very widespread and would require more dedicated studies

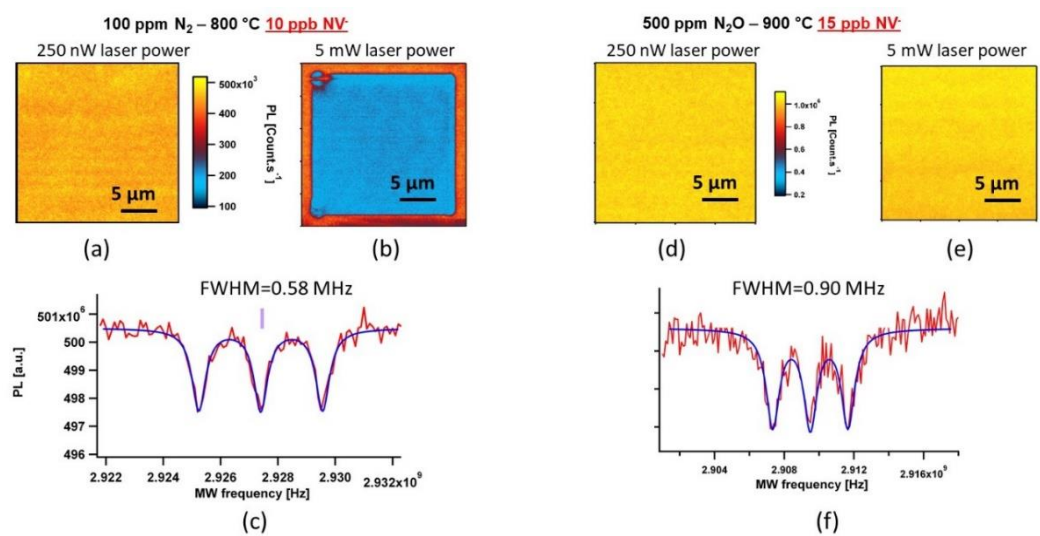

**Figure 5.** (a) and (b) PL images of a CVD diamond film grown with 100 ppm of $N_2$ in the gas phase obtained with a confocal scanning microscope (laser, 532 nm), allowing estimating the NV$^-$ density to 10 ppb and showing the poor photostability under high laser power. (c) ODMR spectra obtained under a magnetic field of 3 mT and where the characteristic hyperfine splitting is observed with a FWHM of 0.58 MHz. (d) and (e) PL images on a CVD diamond sample grown with 500 ppm of $N_2O$ in the gas phase allowing estimating the NV$^-$ density to 15 ppb and showing improved photostability even under high laser power. (f) ODMR spectra obtained under a magnetic field of 3 mT showing a similar hyperfine splitting of 0.9 MHz.

Besides NV, other colour centres can be introduced in-situ in CVD-grown diamonds. However as compared to the HPHT process or to ex-situ implantation, the introduction of a wide variety of impurities is relatively limited. SiV and GeV centres have been obtained through addition of a varying amount of silane or germane gases ($SiH_4$ and $GeH_4$) [149, 150]. For those elements though, doping using a solid-state source (like a small piece of Si or SiC placed near the growing diamond) is usually the preferred approach due to its simplicity and non-toxicity [151, 152]. Dopants that modify the Fermi level of the semi-conducting diamond crystal such as phosphorous (n-type) or boron (p-type) can also be advantageously used to tune the charge state of the colour centres (see part 4.4). This is generally achieved through additions of TBP (tri-butyl phosphine), TMB (tri-methyl boron) or $B_2H_6$ (diborane) [153]. Although it is not directly involved in the creation of specific colour centres, the addition of $O_2$ during growth is also sometimes explored to improve the crystalline quality of the films thereby potentially improving the coherent properties of in situ doped NV centres [154].

### 4.3. Controlling colour centres density

By in-situ doping, NVs can be created in a wide range of doping levels from isolated single centres to ensembles of several 10's of ppb without any post-treatment simply by tuning the added gas concentration during growth. For some nanoscale sensing applications, or for quantum memories [155] the manipulation of single NVs with long coherence times is required. Extremely low amounts of NVs (or even more, no NV at all) are however particularly difficult to achieve and require carefully purified gas sources and leak-tight reactor chambers. While high-purity "electronic grade" commercial diamonds (specified as $N_s$ < 5 ppb) do not normally display luminescence originating from NVs, it has been shown that after annealing at high temperature (1600 °C for 4 h), vacancies diffuse and are able to associate to nitrogen [124]. This leads to up to 1 NV/µm² and indicates that even carefully prepared CVD diamond films may still contain a low but non-negligible residual background of nitrogen. The creation of isolated NVs (0.1-1 ppb) relies on low additions of $N_2$ during growth in general diluted in hydrogen (around 0.1-10 ppm) and thus requires a precise control of the gas phase composition to achieve the desired concentration [139]. Doping efficiency also strongly depends on growth parameters such as power density, temperature and substrate orientation and can thus vary on different set-ups.

On the other hand, high density NV ensembles are desirable for many quantum sensing schemes [156] since the sensitivity of a given sensor will depend on the square root of the number of sensing spins [13] (see part 2, eq. 2). However obtaining very high NV concentrations (> 100 ppb) is problematical by direct CVD growth since large additions of $N_2$, above typically 250 ppm under high power density conditions, is accompanied by a degradation of the surface morphology [157] particularly at the edges of the crystal [158], even leading to a total loss of epitaxy for the highest levels. Nitrogen solubility is also limited in CVD and cannot allow reaching as high levels as those typically reached in type *Ib* HPHT diamonds (a few tens or hundreds of ppm). Under moderate nitrogen doping levels though, CVD diamonds often exhibit a brown colour indicative of the presence of vacancy clusters or dislocations as shown in figure 6(a) [159]. This is due to large absorption features at 510, 360 and 270 nm, the latter being directly correlated with $N_s$ (figure 6(d)). Very high brightness diamond films can be obtained by such doping (figure 6(b)). For thick crystals, $N_s$ concentration can be directly evaluated from the intensity of the 1344 $cm^{-1}$ feature in FTIR as shown in figure 6(c) [160]. It has been proposed that addition of a low amount of oxygen during growth together with nitrogen helps limiting the formation of such defects and the appearance of the brown colour [91]. Highly N-doped CVD diamonds nevertheless contain a large amount of residual defects that strongly reduce their optical properties. HPHT or LPHT post-treatments (> 1500 °C) are often required to improve their colour due to a partial rearrangement or annihilation of point defects [161].

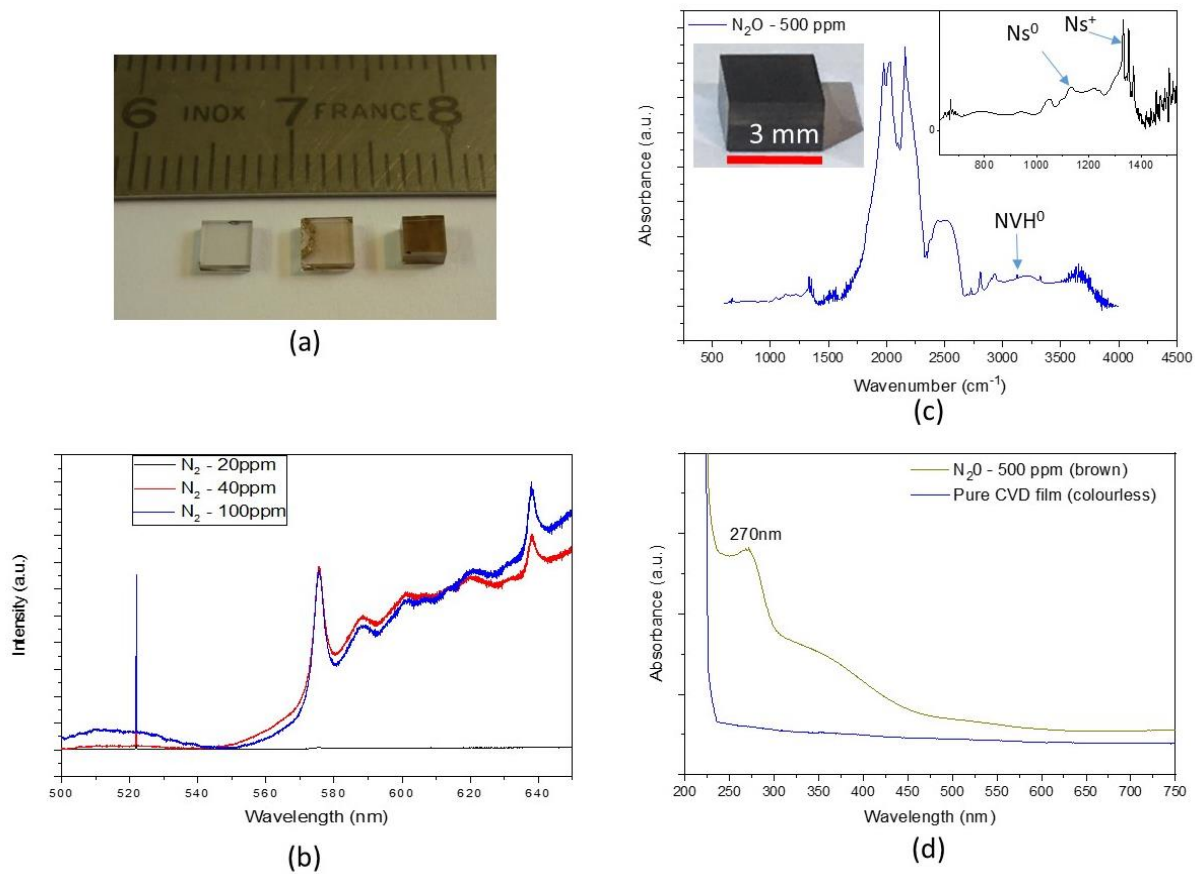

**Figure 6.** (a) Optical images of thick freestanding CVD diamond samples grown with 20 ppm, 40 ppm and 100 ppm of $N_2$ in the gas phase from left to right, ranging from colourless to dark brown. (b) Corresponding Raman and PL spectra performed on these samples showing high emission from $NV^0$ and $NV^-$ luminescence at 575 nm and 637 nm respectively. (c) FTIR spectrum carried out on a thick freestanding CVD diamond sample grown with 500 ppm of $N_2O$ in the gas phase allowing to clearly identify and quantify $N_S^0$, $N_S^+$ and $NVH^0$ defects. (d) UV-Visible absorption spectrum of a high purity CVD diamond plate and a CVD diamond plate grown with 500 ppm of $N_2O$ in the gas phase. Absorption at 270 nm is due to $N_S$ while other broad absorption bands are related to vacancy clusters.

Besides the difficulty in growing CVD crystals in the presence of a high concentration of nitrogen in the gas phase, an additional issue comes from the fact that NVs represent only a small fraction of the total incorporated N content (see part 4.2). The contribution of $^{14}N$ nuclear spin bath on NV spin's dephasing starts to overcome that of natural isotopic $^{13}C$ for concentrations above 0.1 ppm. At 10 ppm total nitrogen, $T_2$ times drop to about 10 µs [50]. Therefore this leads to a trade-off between high NV density and long coherence times. In order to circumvent this, partial conversion of $N_s$ into NVs can be obtained through an appropriate irradiation using high energy electrons [55] or $He^+$ ions (see part 4.1). Using the later, Kleinsasser et al. [162] achieved $NV^-$ densities of the order of 1 ppm which is only 10 fold lower than the highest densities reported in irradiated *Ib* HPHT diamonds [163] while ESR linewidths remained narrow (200 kHz). N to NV conversion rates of the order of 10-20 % are possible through an appropriate irradiation [164]. In this case dipolar interactions between proximal NV centres might dominate the dephasing rather than NV to N coupling.

### 4.4. Controlling NVs charge state

The NV centre possesses neutral and negative charge states with zero phonon line emissions at 575 and 637 nm respectively (see part 2). Both are usually present in nitrogen-doped single crystal diamonds. Quantum sensors exploit the optical properties of the negatively charged NV centres (spin S = 1) and therefore the neutrally charged centres with spin S = ½ are undesirable. They may lead to overlapping of the PL emission due to broad phonon side bands as well as magnetic noise that degrades spin coherence times. It is generally admitted that NV centres acquire their negative charge from nearby electron donors. One obvious candidate for this charge transfer is the substitutional nitrogen that represents a large fraction of the total nitrogen content in non-irradiated nitrogen-doped HPHT and CVD diamonds [145]. For this reason, under a low power excitation in the range 450-610 nm to limit photo-ionization, the steady-state $NV^-$ centre population is typically about 75 % of the total NV amount [165]. However, this value can vary depending on the nitrogen doping level in the diamond crystal. Type *Ib* HPHT diamonds with high nitrogen content (several tens of ppm) may have a higher proportion of $NV^-$ [166, 167]. For example, in the PL spectra of figure 7(a) with an excitation at 532 nm, $NV^0$ emission is almost undetectable as compared to that from $NV^-$ due to the large amount of nitrogen (circa 100 ppm) in this electron irradiated type *Ib* crystal. To the contrary, when too high a proportion of $N_s$ are converted to NVs by irradiation, there may not be a sufficient amount of donors close enough to NVs to provide the necessary electron. In this case, emission from $NV^-$ tends to saturate while that from $NV^0$ increases, which occurs above a certain irradiation dose [168].

Promoting NV's negative charge state can also be achieved with shallower electron donors than nitrogen such as phosphorous or sulphur [125, 169]. Boron which is an acceptor impurity produces an opposite effect by favouring the neutral charge state. Groot-Berning et al. have clearly shown the effect of co-doping in implanted diamond films [170] as illustrated in figure 7(b). Fermi level tuning can also be achieved with in-situ doped CVD diamond films [171] allowing a fine control over NV's charge state. Intentional doping of diamond by phosphorous during CVD growth is however particularly challenging due to the low doping efficiency of this element into the diamond while n-type conductivity is limited by compensating defects and high activation energy (0.6 eV) [172]. Nevertheless, electrical control over NV's charge state (and emission) has been shown with p-i-n junctions and switching from $NV^-$ to $NV^0$ has been obtained by applying a strong bias [173].

Regarding shallow NVs, their charge state also strongly depends on the chemical species present at the diamond's surface (typically hydrogen or oxygen termination, see figure 7(b) and 7(c)). Since hydrogen termination of diamond films induces a 2D hole gas by a surface transfer doping mechanism [174], this termination is not favourable to negatively charged NVs. On the contrary, oxidative etching of the diamond by heating at temperatures around 450 °C in an $O_2$ atmosphere leaves the surface oxygen-terminated and promotes $NV^-$ [175]. Active tuning of the band bending and thus of NV's charge state, through an electrolytic gate electrode [36] or using a Schottky metal contact [176] have been demonstrated leading to on-demand switching of the PL emission. In general, it should be noted that O-terminated diamond surfaces are preferred since they provide a more chemically stable environment for $NV^-$ centres. This termination is easily obtained through either acidic treatment (typically by dipping in boiling $H_2SO_4/HNO_3$ mixtures), by exposure to a soft microwave oxygen plasma or UV-ozone lamp.

Besides those effects intrinsic to the diamond environment and doping, the magnitude of $NV^0$ and $NV^-$ emission in PL also depends on the excitation wavelength of the laser due to different absorption cross-sections [167]. Figure 7(a) shows that the $NV^0/NV^-$ ratio is strongly affected by changing the laser from 532 nm to 473 nm for a diamond irradiated with an electron dose of $1.5 \times 10^{19}$ cm$^{-2}$. It has indeed been shown that excitation in the blue is more favourable to $NV^0$. There is near equal excitation at 514 nm and in this case, the strengths of the zero-phonon lines is a good indicator of the relative concentrations of the two NV charge states [167]. In order to more precisely measure the charge state ratio, decomposing the PL spectrum obtained with a 532 nm excitation has also been proposed as a more straightforward approach [177].

$NV^-$ ionization into $NV^0$ may occur depending on the laser power used and the excitation wavelength. This process leads to blinking issues and limits the maximum spin polarization accessible. Photon energies higher than 2.6 eV are required for direct ionization [165]. However, ionization may proceed through a 2-photon absorption process using a green laser [178]. The first photon induces a transition to the excited state of the defect while the second photon excites the electron to the conduction band of the diamond. This is accompanied by the creation of a photocurrent which can be usefully exploited to electrically detect the spin-state of the NV centre in the so-called PDMR scheme (Photocurrent Detection of Magnetic Resonance) [51, 179]. In a similar way, electron induced ionization of NV centres explains that only emission from the neutral charge state of the defect is detectable in cathodoluminescence [180]. In general the reverse process (recombination) that turns $NV^-$ back into $NV^0$ takes place with a typical time of 500 µs [165] and is promoted by red excitation. Multicolour illumination with a near-Infrared laser increases the $NV^-$ steady-state population with respect to $NV^0$ and strongly improves the spin read-out fidelity [181]. In some cases though, long term ionization can occur with the recombination being inhibited due to the presence of charge traps in the crystal that prevent an efficient diffusion of carriers [182]. Based on a multicolour excitation, optical patterning of the PL on the diamond's surface is then possible allowing foreseeing storage applications [85]. PL extinction for a long period of time is for example illustrated in figure 5(b) under high power green laser [147].

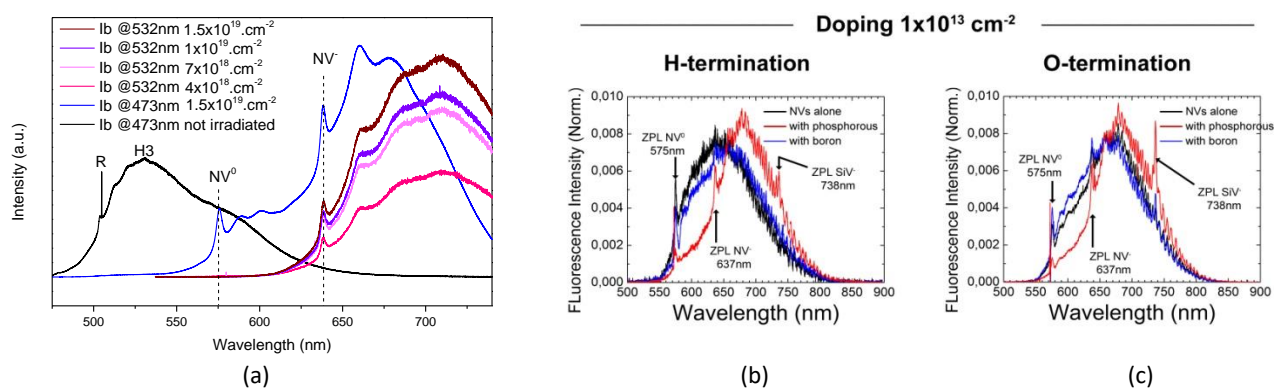

**Figure 7.** (a) PL emission of a type Ib HPHT diamond that has been irradiated by a 1 MeV electron beam at various doses as indicated in the legend (Collaboration ICR, Marseille). Excitation with a 473 or 532 nm laser was carried out. (b) and (c) PL showing the effect of co-doping with boron or phosphorous on the charge state of NV centres for both H-terminated or O-terminated (100) diamond surfaces. (Adapted from ref [170])

## 5. Controlling the spatial localisation of NV centres

Beyond the production of NV-doped diamond layers with good coherence time and stability, for most applications, spatial localization of $NV^-$ defects in the diamond host material is necessary. For instance, in wide-field imaging magnetometry, nanometre-thin layers highly doped with $NV^-$ defects should be located at or slightly below the surface [60, 183] in order to precisely control the distance between the interacting spins and the sample to be measured. Such thin layers are usually called "delta-doped" in analogy with works carried out on electronic doping of semiconductors that achieves high mobility channels with high dopant concentration through such a spatial confinement [184, 185].

The main approach to introduce localized nitrogen atoms into diamond in order to generate $NV^-$ defects is *ex-situ* ion-beam nitrogen implantation followed by annealing which allows for a spatial control of the nitrogen atom's position ultimately limited by ion channelling and straggling effects [121] (see part 4.1.). For nitrogen atoms implanted near the diamond surface with a few keV energy, the depth resolution is typically in the range of few nanometres [183, 186]. However, unwanted paramagnetic defects presumably created during the implantation process reduce the spin coherence time of implanted $NV^-$ defects [183], even if significant improvements can be achieved with optimized irradiation and annealing procedures [187, 188]. For these reasons, directly grown-in NV centres obtained by intentionally adding a nitrogen precursor to the gas phase during diamond growth by CVD [139, 189] is advantageous especially if spatial positioning is achieved. This approach is discussed here.

### 5.1. Positioning NVs in a thin layer ("delta-doping")

Several growth strategies have been developed to localize NV centres in-depth during CVD growth. The simplest way is to turn on and off the $N_2$ input flow which allows growing nitrogen-doped stacked layers as illustrated in figure 8(a). However, in this case, changing the gas phase composition is hampered by long residence times $t_r$ of gas species in the plasma chamber. Indeed, for a typical incoming gas flow of 0.5 litre per minute and a volume for the plasma chamber of a few litres [93], several minutes are required to completely renew the gaseous environment which complicates the growth of stacked layers with sharp transitions [140]. This is all the more difficult since, as previously described, under high microwave powers which are required to efficiently dissociate $N_2$, relatively high growth rates are reached (a few µm/h). Consequently, there is a trade-off between high doping efficiency and high thickness control which involves very low growth rates. Growth techniques with nanometre-scale resolution have already shown promising results in magnetometry [190-192] but they require very low power densities (typically 750 W for a pressure of about 30 mbar) and low methane additions (< 0.5 %) to reduce growth rates down to a few nm/h. This in turn limits achievable NV concentrations to typically $10^8$ to $10^{12}$ cm$^{-3}$. Moreover, at low-power densities, it is difficult to obtain a high crystalline quality especially if one wants to grow the thick buffer layer (> 10 µm) [94, 193] that is required to limit influence of the substrate on the overall luminescence. Nevertheless, delta-doped layers with highly confined NVs located in a $^{12}C$ layer have been produced (see figure 8(b-d)) [141] and have exhibited long coherence times of several hundreds of µs when the surface is sufficiently far away and/or free of defects [194]. This technique has also been coupled to local $^{12}C$ implantation in order to generate 3D profiles of NVs within the diamond sample [195].

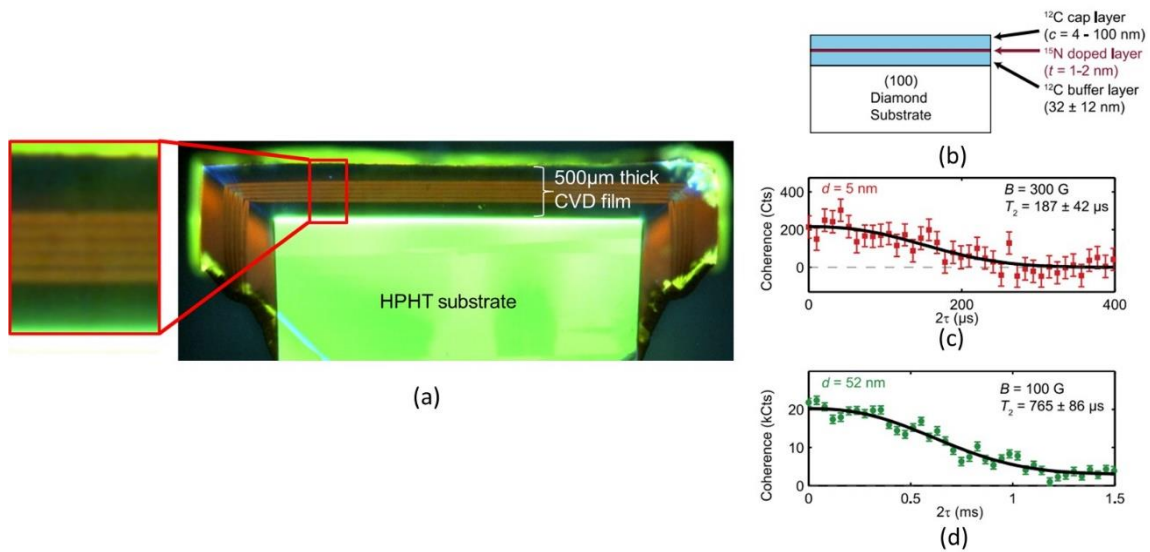

**Figure 8.** (a) Cross section PL image of a 500 µm thick film showing 5 nitrogen-doped layers (in the inset). (b) Schematics showing a delta-doped layer containing NVs (with 15N) and sandwiched between 2 intrinsic layers grown with $^{12}C$. (c) and (d) Hahn echo measurements showing coherence times at different distances from the surface for NV centres located in this delta-doped layer. Adapted from Ohno et al. [141].

An alternative way to localize NV centres in-depth without changing the incoming gas flow is to modify the substrate temperature during growth. Indeed it has been clearly shown that nitrogen doping efficiency strongly depends on temperature [157, 196]. The main advantage of this approach is that a small change of pressure and/or microwave power allows decreasing or increasing temperature by more than one hundred degrees in a matter of seconds especially for high microwave power operating systems. Fast temperature variations can thus be harnessed to create abrupt interfaces that couldn't be easily obtained through gas phase tuning. It is then possible, as illustrated in figure 9, to fabricate nitrogen-doped stacked layers with thicknesses as low as a few hundreds of nanometres at high microwave power densities leading to high N$_2$ dissociation efficiencies and thus high-quality diamond films as well as thick buffer layers.

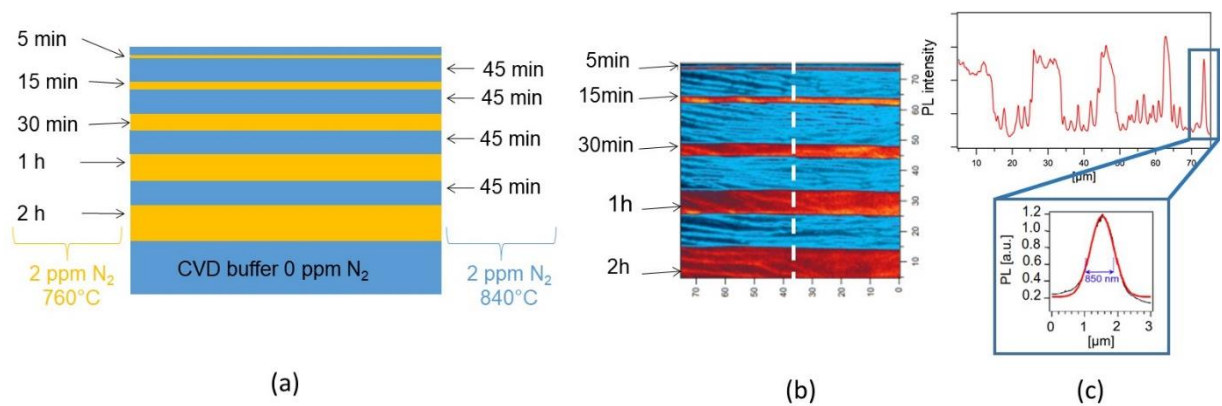

**Figure 9.** (a) Sketch of the cross-section of a CVD multilayer obtained by varying the growth temperature between 760 °C (orange layers) and 840 °C (blue layers) with increasingly short times for the low-temperature step. 2 ppm of N$_2$ were constantly added to the gas phase. (b) PL raster scan of the cross-section of the sample revealing the presence of the 5 low-temperature layers were NV concentration is higher. (c) PL intensity line-cut along the white dashed line shown in (b). The inset shows data fitting with a Gaussian function of the thinnest nitrogen doped layer, leading to a full-width at half maximum of 850 nm.

Another approach to couple the use of high microwave power densities and the possibility of growing delta nitrogen-doped layers is to quickly move the substrate holder in and out of the plasma region, where growth takes place, which seems very easy from a conceptual point of view. Nevertheless, the main difficulty lies in the fact that diamond growth reactors are based on resonant cavities and the substrate holder is part of this cavity. Thus, most of the time the plasma discharge is affected by a displacement of the holder [197]. In addition, to avoid too much a variation of substrate temperature, additional heating systems are sometimes required to compensate for the heating loss from the plasma source [198]. In order to circumvent those issues, substrate holders can be designed so that only a small central part that supports the sample is translated. Reactor designs also exist in which the cavity is relatively unaffected by movement of the holder (such as "egg-style" Aixtron reactors) [199]. This approach thus requires some engineering development and is relatively complicated.

### 5.2. Controlling NV's distance from the surface

As previously detailed, the positioning of NV centres in a thin layer within the diamond crystal is particularly desirable for many sensing applications. In addition, the distance at which NVs are located from the surface also plays a critical role since it will directly affect their ability to sense a magnetic field for instance. In general to maximize the interaction, this distance should be short (i.e. shallow NVs). However the presence of defects or impurities at the surface also strongly affects both the charge state and the coherence time of NV centres due to the induced surface electronic spin bath or magnetic noise induced by surface spins [200]. Shallow NVs have been produced both by direct delta-doping [201] and by low energy ion implantation [186]. Near-surface NVs can also be created by converting incorporated nitrogen through low energy electron irradiation [202].

Precisely measuring NV's distance from the surface is not trivial and requires the development of specific procedures. One possible way is to detect the nuclear magnetic resonance signal from protons in adsorbed species at the surface (such as immersion oil) [203], by coating the surface with an element inducing strong magnetic noise [204] or by approaching a magnetized AFM tip [205]. While deep NVs (depth > 50 nm) in a low concentration nitrogen environment exhibit typical T$_2$ times of 300-500 μs, this figure drops by 1 or 2 orders of magnitude (i.e. 1-10 μs) when the NV to surface distance is reduced to below 5 nm [206]. An overgrowth step in order to keep NV centres away from the surface allows improving Hahn echo $T_2$ time as illustrated on figure 10(a).

Removing defects at the surface is a way to extend coherence times of shallow NVs. An order of magnitude extension has been demonstrated by Sangtawesin et al. [207] through a series of well-controlled surface treatments that aim at recovering a surface state as perfect as possible. This includes polishing down to a roughness below 0.1 nm, removing surface damage by $Ar/Cl_2/O_2$ plasma etching, triacid cleaning (perchloric, nitric and sulfuric acid) followed by oxygen annealing at 450°C to fully oxygen-terminate the surface. Surface termination of the diamond indeed strongly affects shallow NV's properties [208]. Improvements as compared to an as-grown or as-implanted surface have been demonstrated through fluorine termination with an $SF_6$ plasma treatment [209] (see figure 10(b)), nitrogen termination using a high power plasma [210] and oxygen termination as well depending on the initial orientation of the surface, (113) for example as suggested by Li et al. [211]. In addition a soft RIE (Reactive Ion Etching) oxygen plasma etching of the implanted surface, with nanometre-scale precision, has allowed bringing NVs closer to the surface (circa 4 nm) while preserving good coherence properties ($T_2$ up to 30 µs) [212].

Finally, overgrowing a diamond film implanted with a shallow NV pattern provides a way to push further away the surface and can be exploited to improve coherence properties and stabilize NV's negative charge state while still benefiting from the placement accuracy of low energy implantation (i.e. circa 10 nm) [213]. However this approach also comes with additional issues. Impurities such as SiV and defects (dislocations for example) can be preferentially incorporated at the growth interface [214]. Slight etching of the surface as well as passivation of implanted NVs by hydrogen diffusion may also lead to the partial disappearance of the pattern [215, 216].

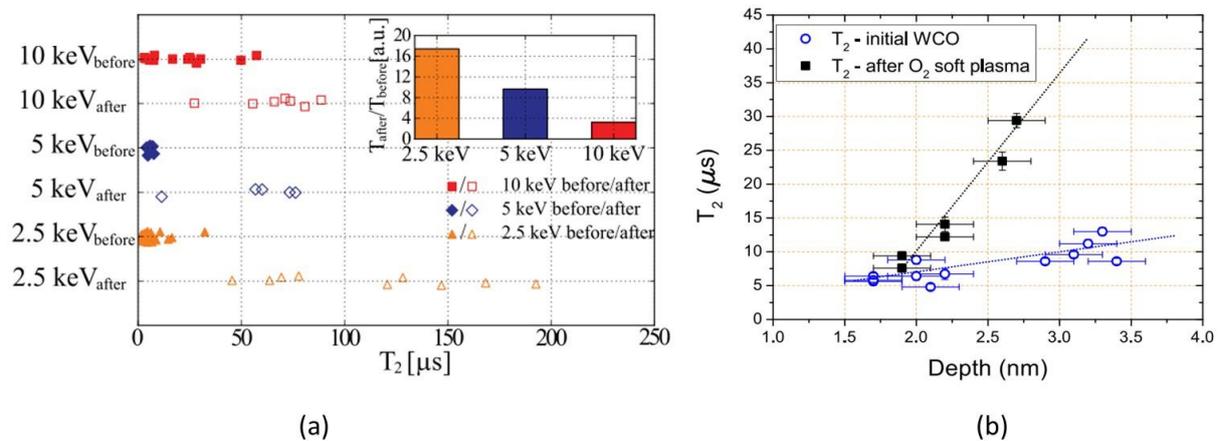

(a)  (b)

**Figure 10.** (a) Increase in the Hahn echo coherence time $T_2$ for single NV centres before and after an overgrowth process. The inset shows the relative increase of the mean coherence times exceeding one order of magnitude for the 2.5 keV implanted NV centres corresponding to the shallowest NV centres. Adapted from Staudacher et al [213]. (b) Evolution of $T_2$ time versus depth showing that an appropriate surface treatment can allow favourably extending coherence time near the surface. Adapted from Favaro de Oliveira et al [212].

### 5.3. Positioning NVs in-plane

It is very challenging to deterministically localize defects in the growing plane (i.e. in the X-Y direction) by CVD. This has been achieved on specific nanostructures either by top-down or bottom-up techniques. In the first case, the pattern created in a hard mask by optical or electron lithography is transferred to a diamond layer containing NV centres (either by growth or implantation) through Reactive Ion Etching (RIE) resulting in colour centres located in pillars as illustrated in figure 11(a). This process has been initially developed in order to obtain high aspect ratio pillared and conical structures that lead to a local field enhancement effect of the emitting surface[217-219]. It has also been successfully applied to nitrogen-doped diamond films in order to create arrays of nanopillars exhibiting NV fluorescence as shown in figure 11(b) [220]. Such pillars can even be usefully exploited to locally perform magnetic sensing when attached to an AFM tip for instance [221] or to be used for sensing at the nanoscale [12]. Beyond the localized NV emission, an additional positive effect of developing such structures is the improved wave-guiding of the NV emission particularly in the case of growth in the [111] direction as shown by Neu *et al* [222]. In this configuration, the preferential orientation of the NV axis in the direction of the pillar (i.e. dipole perpendicular to the pillar axis) leads to an enhanced coupling efficiency and thus extraction of light (see figure 11(c)).

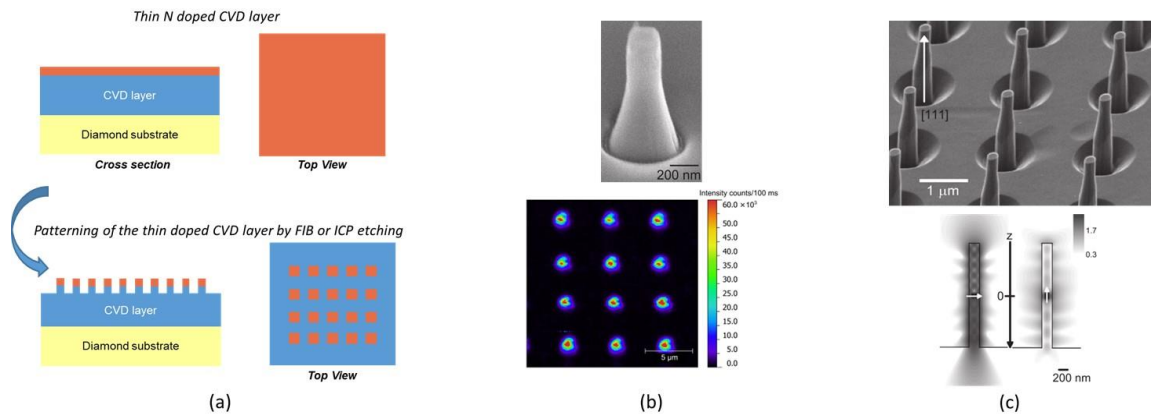

**Figure 11.** (a) Sketch of the fabrication process to create NV centres localized in pillars. (b) Nanopillars obtained by ICP etching with a chromium mask on top of a [100]-oriented CVD diamond single crystal and corresponding micro-PL mapping of an array of such pillars. Adapted from Widmann et al [220]. (c) Nanopillars obtained by ICP etching of a [111]-oriented CVD diamond single crystal and simulations showing wave-guiding of the emission from a single dipole oriented perpendicularly to (left) or parallel to (right) the nanopillar axis. The coupling efficiency is enhanced in the first case [222].

There have been only a few reports on the fabrication of diamond nanostructures based on a bottom-up approach. The formation of nanopillars by plasma etching, described previously, can be followed by an overgrowth step in order to form a thin layer in which impurities (nitrogen or silicon) can be intentionally introduced [223, 224]. Depending on the growth conditions, the shape of those pyramidal features can be varied according to the relative growth rates of different crystalline planes. This additional step can thus increase the concentration and localization of active colour centres [225-227] which can be useful to create quantum sensors or single photon sources based on this material [228]. In a similar way, fabrication of p-n electronic devices has been achieved by locally growing phosphorous doped layers on patterned diamond single crystals [229].

An uncommon way to fabricate arrays of NV centres in CVD diamond is based on a 2-step approach illustrated in figure 12(a). First, a pattern of micro-holes is created on a (100)-oriented high purity diamond substrate by optical lithography followed by RIE etching. The lateral orientation of the holes is chosen along the (110) directions. In a second step, overgrowth is performed by CVD so that the created holes were "re-filled" with nitrogen-doped diamond with limited growth on the top surface. This implies choosing growth conditions leading to a high (111) to (100) growth rate ratio, i.e. low methane concentrations and high substrate temperatures [230, 231]. The hole array totally disappears leading to a smooth surface (figure 12(b)). NV centre distribution can be assessed by observing the patterned region by CL. Figure 12(c) shows the CL image acquired at a wavelength of 575 nm, which corresponds to emission from NV$^0$ centres in diamond. The pattern of holes is revealed by the presence of localized NV centres.

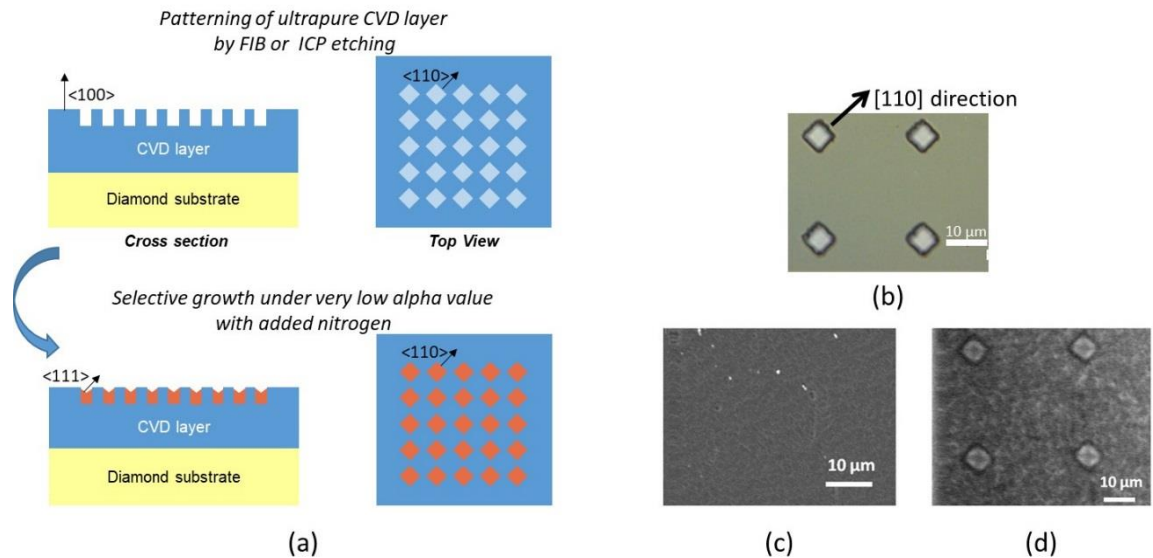

**Figure 12.** (a) Sketch of the fabrication process allowing obtaining arrays of NV centres localized at the surface of CVD diamond. (b) Optical picture of the top CVD layer after ICP etching where square holes are visible. (c) Secondary electron image of the patterned area after CVD growth which shows a smooth surface with no visible holes remaining. (d) CL image at 575 nm of the same region where bright luminescence from NV centres is visible in the "re-filled" holes.

## 6. Controlling NV orientation

NV centres' quantization axis is oriented along the <111> crystallographic directions and can thus have 4 equivalent directions in the crystal. In order to improve either the sensitivity or the easiness of use of a magnetic sensor working with an ensemble of NV centres, promoting a specific orientation among those 4 is a great advantage. CVD growth on substrates with alternative orientations provides a way to do that [232]. In general the most out-of-plane crystallographic directions of NV centres are more likely to show-up following CVD growth. When growth is performed on standard (100)-oriented substrates, all 4 directions having an equivalent angle with respect to the surface, there is an equiprobable formation of the four possible NV orientations (see figure 13(a)). Thus a maximum of 25 % of the total number of NV centres will be oriented with the desired angle. This has motivated research to grow on other non-classical growth orientations.

### 6.1. Growth on [110]-oriented substrates

The first report of preferential alignment for grown-in NV centres was done using CVD-grown [110]-oriented films [143]. On this orientation, $[\bar{1}1\bar{1}]$ and $[1\bar{1}\bar{1}]$ directions are in the (110) plane while [111] and $[\bar{1}\bar{1}1]$ are out-of-plane making a similar angle of 35.2° (see figure 13(b)). This particular configuration thus leads to 2 main populations among the 4 possible directions (50 % preferential orientation). Moreover, the specific configuration of the atomic structure on the plane suggests that defects are incorporated as a unit of a nitrogen and a vacancy rather than formed later by diffusion of a nearby vacancy [143]. Despite this preferential alignment, the (110)-orientation has not been the subject of many studies or applications [233-235]. In fact (110)-oriented substrates are commercially available but with a limited size (typically 3x3 mm² or below). Due to cuboctahedral growth, when [110]-oriented diamond plates are extracted from the HPHT-grown crystal, multiple sectors are usually obtained [232] which constitutes an important drawback since sector boundaries lead to stress and defect formation [94, 236]. In addition, since homoepitaxial layers on (110)-oriented diamond substrates have one of the highest growth rates under standard CVD conditions [230], the top surface tend to rapidly disappear during growth limiting the final surface area when thick films are required. This point is illustrated in figure 14(a) and 14(b) in which the surface area decreases by a factor of 2 after a 30 h of growth even if 300 μm-thick films with high-quality can be grown at 10 μm/h.

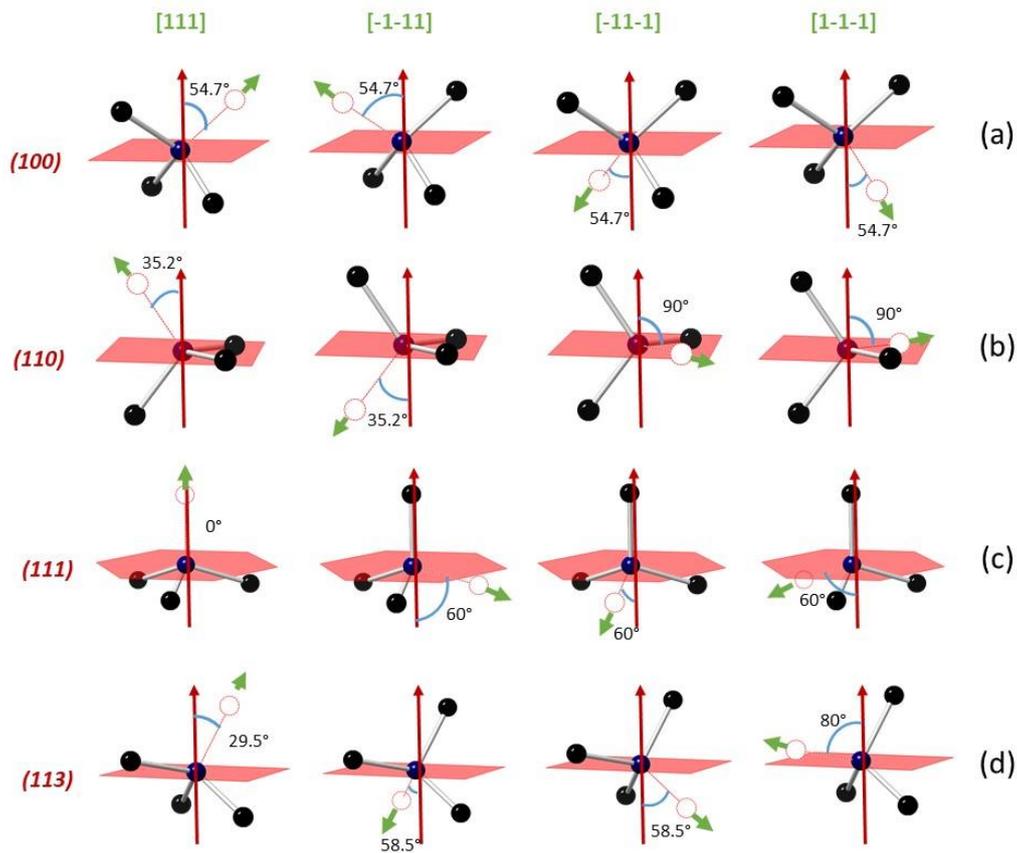

**Figure 13.** Schematic drawing of the four possible NV defect orientations (in green) for a (a) [100]-oriented diamond sample, (b) [110]-oriented diamond sample, (c) [111]-oriented diamond sample and (d) [113]-oriented diamond sample. The most out-of-plane directions (lowest angles) have the highest probability for preferential orientation during CVD growth. (Black: carbon, blue: nitrogen, red: vacancy).

In order to avoid the use of (110)-oriented substrates, it has been reported that growth on a (100) surface onto which step bunching occurs can also lead to specific alignment of NVs [143]. This is due to the fact that step edges mainly consist of (110) planes [237] (figure 14(c)). Therefore preferential alignment of NVs on the risers of those steps is observed. Unfortunately, in this case, the NVs uniformity is very poor since they are mainly localized in stripes corresponding to the displacement of step edges during growth [238] (figure 14(d)). For all these reasons, (110) has been supplanted by other orientations that are more advantageous in terms of preferential orientation and/or easiness of growth.

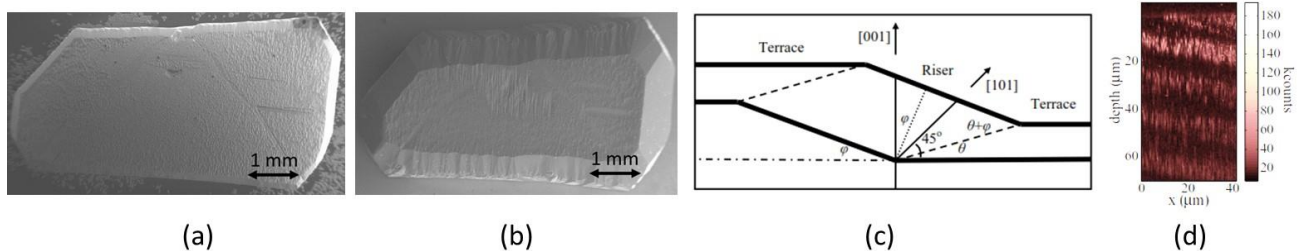

**Figure 14.** Surface morphology of a (110)-oriented CVD layer after a growth of (a) 6 hours (60 μm) and (b) 30 hours (300 μm). The top surface area has eventually decreased by a factor of 2. (c) Scheme of macro-steps formed at the surface of a [100]-oriented diamond film during growth where risers and terraces are identified. This image is adapted from Martineau et al. [237]. (d) Preferential incorporation of NVs at the risers of the steps with (110) orientation. This image is adapted from Pham et al. [238].

## 6.2. Growth on [111]-oriented substrates

If we consider a [111]-oriented sample, one NV direction will be perpendicular to the top surface (0° angle) while the three others will make a similar angle of 60° with respect to this direction (See figure 13(c)). Therefore, preferential alignment of NVs along this perpendicular direction will occur during growth [239-241]. Using an atomistic layer-by-layer growth model, Miyazaki et al. have shown that incorporation of nitrogen at kinks during the displacement of [$\bar{1}\bar{1}2$] step edges is likely to lead to such preferential orientation [242]. In addition, it has been shown that, among this family, alignment of NVs along [111] rather [$\bar{1}\bar{1}\bar{1}$] (i.e. N-V rather than V-N) has a higher probability to occur [241].

[111]-oriented HPHT diamond crystals can be purchased although their availability and size (typically 2×2 mm²) are quite limited. (111) plates can be extracted by slicing large cuboctahedral HPHT crystals (or CVD stones) with a 54.7° angle from the top (100) sector. Alternatively under certain HPHT conditions, octahedral growth is possible leading to pyramidal diamonds with <111>-oriented side facets [79] from which the fabrication of triangular (111) plates with minimal material loss and larger growth sector is possible. This crystalline orientation being particularly hard mechanically, it is extremely difficult to polish along an exact [111] direction using a standard *scaife* technique [243, 244]. A slight misorientation angle (1 to 2 °) is usually needed to obtain low roughness and reasonable polishing rates. As a consequence most as-received (111) substrates are not exactly oriented. Alternative polishing techniques have been proposed that make use of chemical agents or UV light assistance [245, 246] to obtain a better surface finish.

The growth of CVD diamond on this specific orientation is known to be very difficult due to twinning, defects formation or impurity incorporation [247-249] which usually leads to lower carrier mobilities as compared to a conventional [100]-orientation [250]. Nevertheless, with the use of (111) substrates having a controlled misorientation of 2° along the [$\bar{1}\bar{1}2$] direction, atomically-flat thin films compatible with high-quality electronic devices have already been produced [251-253]. For thicker films, the probability for defect formation is higher due to the low energy difference between normal and twinned configurations as described by Butler et al. [254]. However specific growth conditions can be chosen so that a formed penetration twin is quickly overgrown by its parent face and cannot develop further [231]. This corresponds to an $\alpha$ value below 1.5; with $\alpha = \sqrt{3}\frac{V_{(100)}}{V_{(111)}}$ and where $v_{(100)}$ and $v_{(111)}$ are the growth rates of the (100) and (111) planes respectively. This is illustrated in figure 15(a) and 15(b) for 100 µm-thick intrinsic diamond films for which the use of high temperatures (> 1000°C) and low methane concentrations (< 1 %) allows minimizing $\alpha$ and keeping smooth morphologies. A relatively high crystalline quality as judged by Raman/PL has been demonstrated with growth rates as high as 6 µm/h which makes this process viable for the growth of thicker material. More information can be found in reference [80].

When a small amount of nitrogen is added to the gas phase (a few ppm), NV centres aligned along the [111] direction, i.e. perpendicularly to the surface, are preferentially formed [239-241] with a probability from 94 to 100 % (see figure 15(c) and 15(e)). This orientation further facilitates their integration into devices by optimizing the alignment with respect to an applied magnetic field and maximizing their emission when embedded in nanostructures [222]. Furthermore, if growth is performed using a $^{13}$C-depleted carbon source the coherence time can be enhanced to the ms range as illustrated in figure 15(f) for single isolated centres (see figure 15(d) and part 4.2). Obtaining high density NV ensembles is however rather difficult and may lead to a partial loss of orientation. Indeed, the optimal growth conditions window allowing keeping an α parameter lower than 1.5 strongly reduces when nitrogen concentration is increased in the feed gas. Nevertheless by using specific growth conditions in a low plasma density CVD, several tens of ppb of NV centres have recently been obtained with good alignment and moderate $T_2$ times of the order of a few µs [194, 255]. In particular it has been shown that lower growth temperatures (circa 800 °C rather than 1000 °C) are preferable to promote preferential alignment of those ensembles. This is consistent with the fact that at high annealing temperatures above 1000 °C, a partial dissociation of NVs accompanied by vacancy migration might be the reason for re-orientation [256]. The films' thicknesses remained very limited though due the induced stress and low growth rates involved which currently limits the use of such CVD layers. For those reasons the use of alternative orientations to (111) have been considered in order to benefit from preferential orientation as well as higher tunability of NV density in thicker layers.

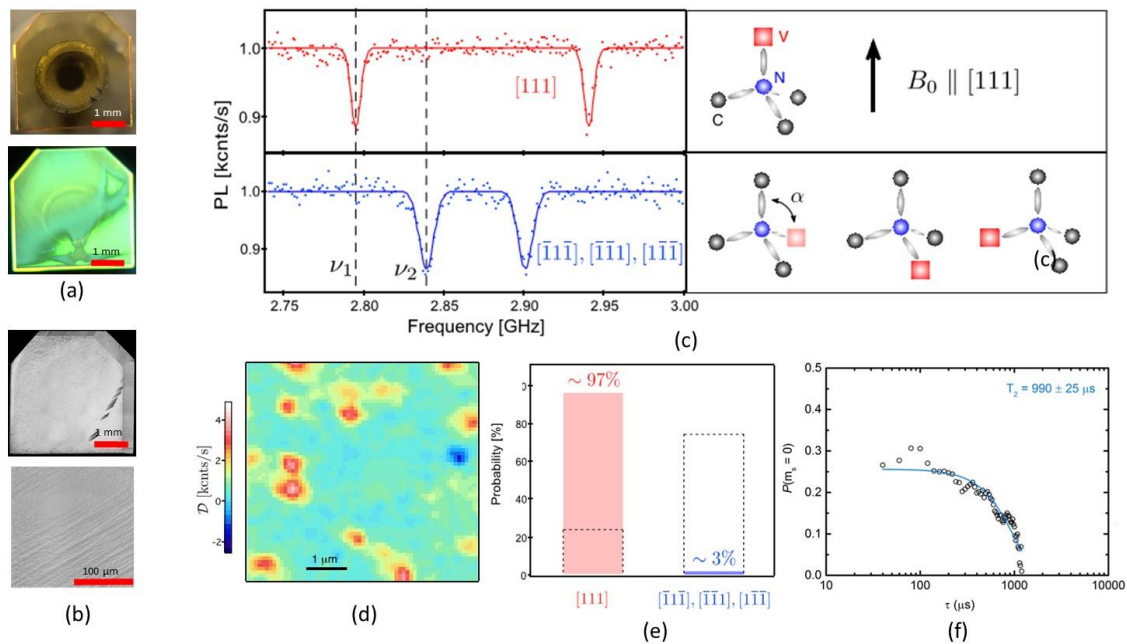

**Figure 15.** (a) Optical and PL images of a [111]-oriented HPHT substrate obtained from a cuboctahedral diamond (Collaboration with the Sobolev Institute of Geology and Mineralogy). (b) Laser microscope images of a 100 μm-thick CVD layer grown on this substrate under optimized conditions. (c) ODMR signal of the 4 NV orientations with magnetic field aligned along [111]. (d) Confocal optical microscopy image of isolated NV centres oriented along [111] in red colour. (e) 97% preferential orientation has been estimated from analysis of more than 200 centres. (f) Spin echo signal measured for NV centres in a $^{13}$C depleted (111) diamond leading to almost 1 ms coherence time. (Collaboration University of Basel).

### 6.3. Growth on [113]-oriented substrates

The choice of this alternative orientation has been guided by the fact that in this configuration, the $[1\bar{1}\bar{1}]$ direction is almost in the (113) plane (80° angle with respect to the [113] direction) while the [111] direction is the most out-of-plane with an upward angle of 29.5°. The $[\bar{1}1\bar{1}]$ and $[\bar{1}\bar{1}1]$ directions make a similar downward angle of 58.5° with the [113] direction and are thus closer to the plane (See figure 13(d)). One of the limitations to the use of [113]-oriented single crystal diamond substrates is that they are not currently commercially available and need to be specifically fabricated on demand. Starting from relatively thick HPHT or CVD crystals with six {100} faces, (113) planes can be prepared by laser cutting and polishing the top face with an angle of 25.2° (± 0.5°) towards the [110] direction. For example in Ref. [257], cylindrical diamond substrates with a 2 mm-diameter have been produced using this procedure. (113) crystallographic faces are stable under certain CVD growth conditions [258, 259] and this orientation is suitable for thick layer growth. This stability is likely related to the fact that the (113) plane undergoes a surface reconstruction in the presence of a hydrogen plasma in a similar fashion to what has been reported for silicon, thus decreasing the surface energy on this orientation[260, 261]. Under high plasma density conditions 460 μm-thick CVD diamond films with smooth morphologies and free of non-epitaxial features have been grown (figure 16(a)). The corresponding growth rate of 15 μm/h was higher by a factor of 2 than that obtained on (100) under the same conditions. A decrease of the available surface area with thickness is noticeable although it remains limited and compatible with the synthesis of mm-thick films. It is essential to note that growth temperature range (700 to 1000 °C) as well as $N_2$ addition of up to a few tens of ppm in the gas phase can be used which is not as restrictive as for (111) growth, as illustrated in figure 16(b-c). PL images (2nd row of figure 16(a-c)) shows only limited blue fluorescence related to the presence of dislocations (band A) [262] and indicates a limited stress in contrast with growth on (111). Green emission originates from the HPHT seed. A weak orange luminescence related to NVs centres begins to appear for the highest doping level which was confirmed by CL performed at 110 K (figure 16(d)). NV defect orientation was then experimentally measured by recording ESR spectra while applying a static magnetic field along the [113] direction. Only three of the 4 possible orientations were detected with 2 of them making a similar angle with the field, leading to 2 different ESR spectra presented in figure 16(e). The probability of occurrence of each NV defect orientations was estimated by recording ESR spectra over a set of about 200 single NV defects. The resulting statistical distribution is shown in figure 16(f) and a preferential orientation of 73% was clearly evidenced. This is again consistent with the fact that "in-plane" directions are very unfavourable for NVs creation [143, 238-241] and out-of-plane ones are promoted.

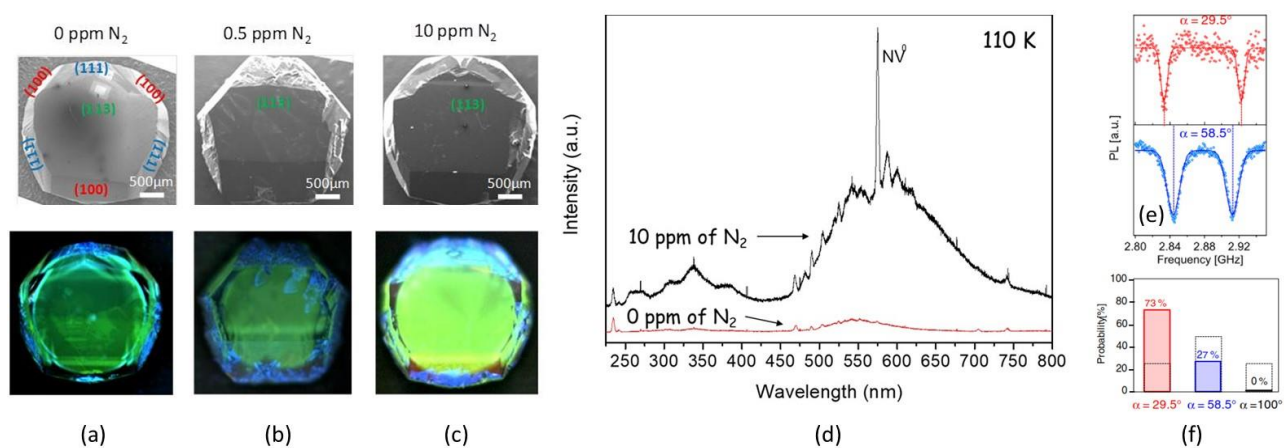

**Figure 16.** (a)–(c) CVD diamond layers grown on [113]-oriented HPHT substrates. (a) non-intentional $N_2$ addition, (b) 0.5 ppm $N_2$, (c) 10 ppm $N_2$, in the gas phase. First row: SEM images where the different crystalline faces can be identified; Second row: PL images recorded with the DiamondView™ equipment under UV light excitation. Green luminescence comes from the HPHT substrate and blue and red luminescence from dislocations/stress and nitrogen related centres respectively in the CVD layer. (d) Cathodoluminescence spectrum recorded at 110 K on CVD layers grown with 0 and 10 ppm of $N_2$ in the gas phase. (e) Orientation-dependent ESR spectra recorded from single NV centres while applying a static magnetic field B=18 G perpendicular to (113) diamond surface plane. The most-in plane $[\bar{1}1\bar{1}]$ direction is not detected. (f) Statistical distribution of NV defect orientations extracted from ESR measurements for a set of about 200 single NV defects. The black dashed lines indicate the expected distribution for randomly oriented NV defects.

As a large range of temperatures is accessible for (113) growth, the influence of this parameter has been assessed [263]. A stacked multilayer alternating high (1000°C) and low (800°C) temperatures has been produced following the scheme illustrated in figure 17(a), leading to a modulation of the nitrogen doping efficiency (see also part 5.1). It has however been found that preferential orientation changes from 50 % at high temperature to a maximum 80 % at low temperature (figure 17(b-c)). This approach thus allows combining both higher doping efficiency with preferential orientation. The spin coherence time ($T_2$) of the NV ensemble has also been measured by applying a Hahn echo sequence. As shown in figure 17(d), the spin echo signal exhibits characteristic collapses and revivals induced by the interaction with a bath of $^{13}C$ nuclear spin[264]. The decay of the envelope leads to a coherence time $T_2$ of 232 ± 6 µs, which is similar to the one commonly obtained with conventional (100) crystals of identical isotopic purity [265]. Therefore the (113) orientation allows having a high degree of engineering of NV defects orientation and density. Although, as compared to (111), NVs are inclined with respect to the surface and only partially oriented, this orientation offers a good compromise particularly for thick CVD diamond layers.

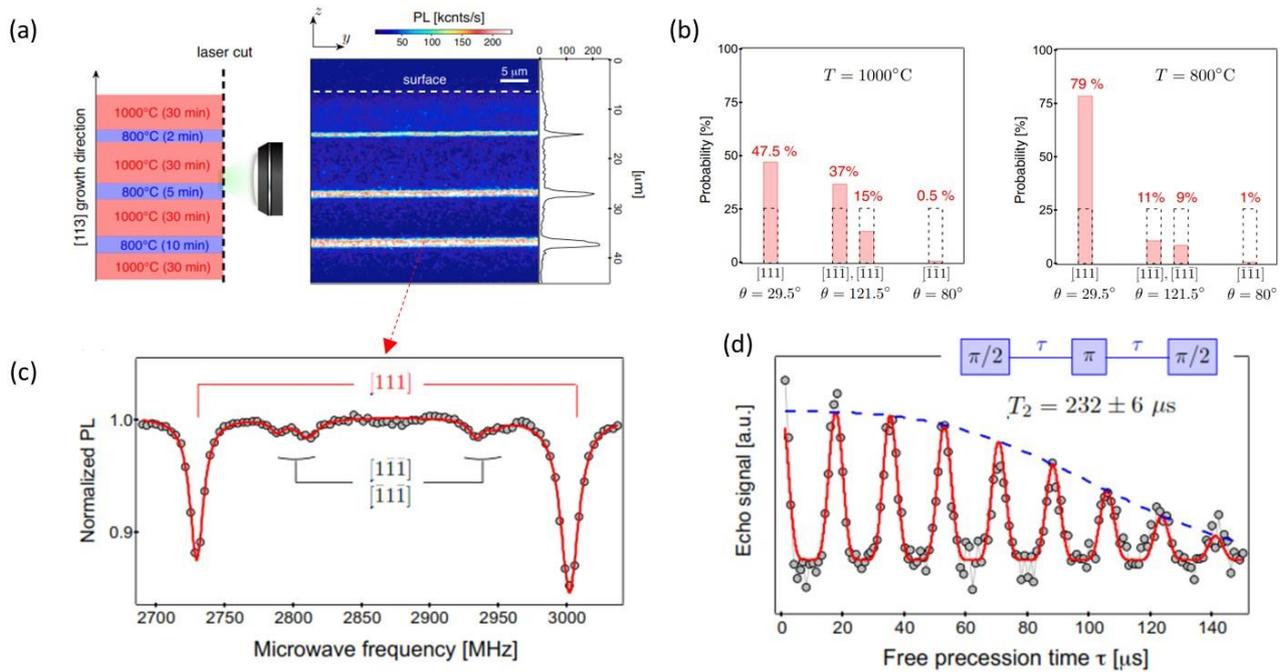

**Figure 17.** (a) Left – Sketch of a sample grown at different temperatures and deposition times on a [113]-oriented diamond substrate. Right – Confocal PL raster scan of a cross-section of the sample obtained by laser cutting. (b) Probability distribution of the NV defect orientation sorted by polar angle θ in samples grown at 1000°C and 800°C. The black dotted bars indicate the expected distribution for a diamond sample without preferential alignment. (c) ESR spectrum of a NV-doped layer grown on a [113]-oriented substrate while applying a static magnetic field which separates all NV defect orientations. (d) Spin-echo signal recorded for the subset of [111]-oriented NV defects with a static magnetic field of ~50 G applied along the [111] axis. The inset shows the spin-echo sequence consisting of resonant microwave (π/2) and (π) pulses separated by a variable free precession time τ.

## 7. Conclusions

The development of single crystal diamond synthesis by microwave-assisted plasma CVD has witnessed considerable progresses over the past 15 years to become nowadays a mature technology. Extensive efforts have been dedicated to the development of high purity diamond films for power electronics through the reduction of residual background impurities down to levels as low as 0.1 ppb for nitrogen and boron. To this end, high quality single crystals have been reliably fabricated with thickness of a few hundreds of µm on various orientations and doping concentrations. Quantum technologies that make use of colour centres in diamond have leveraged on those achievements to unleash the potential of this material system. While HPHT synthesis produces bulk crystals with high crystalline perfection, purity usually remains limited and the technique does not allow for a precise engineering of stacked "quantum grade" layers. On the other hand, CVD has become a key enabling technology for quantum sensing due to its unrivalled control on impurity content and isotopic ratio. The route that is most widely followed is thus to grow a thin active film by CVD with optimised properties for sensing onto a HPHT diamond substrate possessing appropriate quality and orientation.

In particular precisely controlling NVs density in the crystal as well as their close environment is crucial to achieving optimal spin properties for quantum sensing applications. Many studies have thus been dedicated to optimizing the performance of CVD diamonds through a proper engineering of the material. Reducing nearby trapping centres created during growth by choosing alternative doping sources or growth conditions for example, may help stabilizing the charge state of NVs and avoid photobleaching issues. Isotopic control of the $^{13}C/^{12}C$ ratio associated to modification of the Fermi level has also allowed reaching record spin coherence times over 2 ms at room temperature. Beyond that, the coupling of a NV spin to a nearby long-lived nuclear spin has been proposed to further extend quantum storage times or create arrays of entangled systems. It is clear that many sensing applications (such as magnetic resonance imaging) will strongly rely on the future development of such complex coupling schemes and much efforts will need to be dedicated to this research area. To this end, nanometre-scale localisation of NV centres (and other nuclear spins) within the diamond lattice definitely remains an important challenge. So called delta-doped diamond layers have been produced with confinement of the colour centres in very narrow regions at or slightly below the surface. However ex-situ ion implantation or laser irradiation provides much better positioning accuracy than in-situ doped NVs despite slightly worse coherent properties. These approaches will need to be further developed.

The synthesis of thick synthetic diamonds possessing an extremely high amount of NV centres and reasonably good coherence properties is another short-term target that will help pushing further the accuracy and performance of quantum sensors. Producing such crystals is not trivial and requires that adapted growth conditions are found to allow for high doping efficiency while preserving crystalline quality. In many cases, NV/$N_s$ ratio needs to be improved through appropriate irradiation and annealing procedures. One specific asset the CVD technique can build on, is the ability to grow on different substrate orientations leading to partial or even full orientation of NV centres along a specific crystalline direction. The combination of different advantageous properties such as a high NV yield with specific orientation and possibly localisation in the diamond are likely to play a key role in the adoption of this material system.

Nevertheless, efforts need to be pursued to improve the availability of this specially designed material to a broader community. Indeed, only few companies or academic laboratories currently have the know-how to engineer such quantum-grade material. Moreover, the size of the synthetic crystals remains limited to few mm² and it is important to push forward the HPHT and CVD techniques in order to increase the available area. This point is particularly important when micro-fabrication steps are required as it is the case for diamond tips fabrication. The last point that could be highlighted is the importance of defining standards in the production of quantum grade samples in terms of coherence time, NVs concentration, orientation which could allow opening the way of a more reproducible industrial production.

Acknowledgment:
This project has received funding from the European Union's research and innovation program through the Project ASTERIQS under grant agreement No 820394 and through the QUANTERA project MICROSENS n° ANR-18-QUAN-0008-02. It has been also supported by Region Ile-de-France in the framework of DIM SIRTEQ. ANR (Agence Nationale de la Recherche) and CGI (Commissariat à l'Investissement d'Avenir) are also gratefully acknowledged for their financial support through Labex SEAM (Science and Engineering for Advanced Materials and devices), ANR-10-LABX-096 and ANR-18-IDEX-0001, the Diamond-NMR project n° ANR-19-CE29-0017-04 and the SADAHPT project n° ANR-19-CE30-0027-02.